\documentstyle[12pt]{article}
\newcommand{\beq}{\begin{equation}}
\newcommand{\eeq}{\end{equation}}

\def\half{{\textstyle{1\over2}}}

\def\p1half{{\textstyle{{{p+1}\over{2}}}}}

\def\23phalf{{\textstyle{{{23-p}\over{2}}}}}

\begin{document}
\thispagestyle{empty}
\begin{titlepage}

\bigskip
\hskip 3.7in{\vbox{\baselineskip12pt
%\hbox{hep-th/0505xxx}
}}

\bigskip\bigskip
\centerline{\large\bf Thermal Duality and the String Canonical Ensemble}

\bigskip\bigskip
\bigskip\bigskip
\centerline{\bf Shyamoli Chaudhuri \footnote{Email:
shyamolic@yahoo.com} } \centerline{1312 Oak Drive}
\centerline{Blacksburg, VA 24060}
\date{\today}

\bigskip
\begin{abstract}
\vskip 0.1in We derive the free energies of 
both the closed heterotic, and the unoriented, open and closed, type I 
string ensembles, 
consistent with the thermal (Euclidean T-duality) transformations 
on the String/M Duality Web. A crucial role is played by a temperature
dependent Wilson line wrapping Euclidean time, responsible 
for the spontaneous 
breaking of 
supersymmetry at finite temperature while eliminating thermal tachyons,
and determined uniquely by thermal duality.
Conversely, we can show that the absence of a Yang-Mills gauge sector precludes
the possibility of an equilibrium type II canonical ensemble 
prior to the introduction of background Dbranes or fluxes. 
As a consistency check, we verify that our results for the string 
free energy always
 reproduce the $T^{10}$ growth expected 
in the low energy field theory limits while displaying a dramatically slower 
$T^{2}$ growth at temperatures above the string scale.
We present both the low and high temperature expansions for the 
one-loop heterotic and type I string free energies, results which follow from 
an explicit term-by-term evaluation of the modular integrals in the string
mass level expansion. 
\end{abstract}
\end{titlepage}

\section{Introduction}

\vskip 0.1in T-duality invariance, the result of interchanging
small radius with large radius, $R$ $\to$ $\alpha'/R$, is a {\em
spontaneously} broken symmetry in String/M Theory: in other words,
a T-duality transformation on an embedding target space coordinate
will, in general, map a given background of String/M theory to a
different background of the same theory. To be specific, the
circle-compactified $E_8$$\times$$E_8$ heterotic oriented closed
string theory is mapped under a T-duality to the
circle-compactified ${\rm Spin}(32)/{\rm Z}_2$ heterotic oriented
closed string theory \cite{ky,nair,busch,gv,itoyama,ginine}. The
circle-compactified type IIA oriented closed string theory is
mapped to the circle-compactified type IIB oriented closed string
theory \cite{din1,dlp}. And, finally, the unoriented type IB open
and closed string theory is mapped under a T-duality
transformation to the, rather unusual, type I$^{\prime}$
unoriented open and closed string theory: the integer-moded open
string momentum modes associated with the compact coordinate are
mapped to the integer-moded closed string winding modes in the
T-dual type I$^{\prime}$ theory \cite{dlp,dbrane,polchinskibook}.

\vskip 0.1in The results in this paper are a beautiful illustration of the
significance of the Euclidean T-duality transformations linking the six different 
weakly coupled string theory limits of the String/M Duality web 
{\em at finite temperature}.\footnote{The 
idea of pursuing the significance of Euclidean T-duality transformations 
on the different string theory limits of the String/M Duality Web came to me after 
reading Polchinski's discussion of thermal duality in the closed bosonic 
string theory given in his textbook \cite{polchinskibook}, an approach I first considered in 
\cite{decon}. The observation that the one-loop string free energy has a first principles 
derivation from the Polyakov path integral, therefore expressible as an integral over
the fundamental domain of the modular group, is apparent 
in Polchinski's analysis of the closed bosonic
string ensemble in \cite{poltorus}; I simply applied that observation to the six 
supersymmetric string theories. For the heterotic string, I discovered that
some of the necessary inputs for the one-loop vacuum amplitude already existed in the literature 
\cite{ginine,itoyama,gv}, except that their relationship to finite temperature
string theory had not been noted. My presentation
of the type I and type I$^{\prime}$ strings in \cite{decon} was guided by 
the well-established isomorphism to the heterotic strings \cite{polwit}, 
motivated also by the implications for the physics of the 
low energy finite temperature 
gauge theory limit \cite{pair,flux}. Finally, the 
impossibility of
a type II canonical ensemble became apparent to me 
in the follow-up work \cite{fermionic}, where I also discovered 
the significance 
of the temperature dependent Wilson line in both heterotic
and type I ensembles, the connection to conjectures for
strong-weak coupling duality without supersymmetry \cite{bd},
and the $T^2$ growth of the string free energy at high temperatures. 
The realization that the conventional 
lore of the Hagedorn transition is flawed came 
to me in \cite{fermionic,bosonic}, and was explored further in 
subsequent works \cite{holo}. Thus, the presentation that follows in this paper has 
been determined by 
pedagogy, rather than the chronology of my own understanding.}
It is clear that a Wick rotation on $X^0$ maps the
noncompact $SO(9,1)$ Lorentz invariant background of a given
supersymmetric string theory to the corresponding $SO(10)$
invariant background, with an embedding time coordinate of
Euclidean signature. The Wick-rotated $SO(10)$ invariant
background arises naturally in any formulation of equilibrium
string statistical mechanics in the canonical ensemble, the
statistical ensemble characterized by fixed temperature and fixed
spatial volume $(\beta ,V)$. The Polyakov path integral over
connected world-surfaces is formulated in a target spacetime of
fixed spacetime volume. Thus, the one-loop vacuum functional in
the $SO(10)$ invariant background computes precisely the sum over
connected one-loop vacuum graphs in the target space $R^9$ in the
finite temperature vacuum at temperature $T$ \cite{poltorus}. The
appearance of a {\em tachyonic} mode in the string thermal
spectrum is an indication that the worldsheet conformal field
theory is no longer at a fixed point of the 2d Renormalization
Group (RG): the tachyon indicates a relevant flow of the 2d RG.
The question of significance is then as follows: does the relevant
flow terminate in a new infrared fixed point? If so, the new fixed
point determines the true thermal string vacuum. An equilibrium
statistical mechanics of strings requires that this fixed point
belong to a {\em fixed line} parameterized by inverse temperature
$\beta$ \cite{cs}: the precise analog under Wick rotation
of the line of fixed points parameterized by the radius of a
compact spatial coordinate in the $SO(9,1)$ vacuum.

\vskip 0.1in Target spacetime supersymmetry, and its spontaneous
breaking in the thermal vacuum along the line of fixed points
parameterized by $\beta$, introduces new features into this
discussion. We must require compatibility with the expected
properties of the low energy field theory limit where the
contribution from massive string modes has been suppressed,
namely, those of a 10D finite temperature supersymmetric gauge
theory. We must also require consistency with string theoretic
symmetries of both worldsheet, and target space, origin in the
Wick rotated $SO(10)$ invariant background. In particular,
Euclidean T-duality transformations must link the thermal vacua of
the six different supersymmetric string theories in pairs:
heterotic $E_8$$\times$$E_8$ and ${\rm Spin}(32)/{\rm Z}_2$, type
IIA and type IIB, and type IB and type I$^{\prime}$. Remarkably,
we will find as a direct consequence of the T-duality
transformations, that the tachyonic thermal instabilities arising
in all previous attempts to formulate an equilibrium
supersymmetric string statistical mechanics in the canonical
ensemble for the heterotic and type I strings are simply {\em absent}.
For the type II superstrings, in the absence of Dbranes or fluxes, 
we will find that there is no canonical ensemble: the thermal 
tachyon free 
background also has target space supersymmetry, 
as a consequence of modular
invariance.

\vskip 0.1in Having clarified that the canonical ensemble of
heterotic and type I strings is well-defined at all temperatures,
including the temperature regime far above the string mass scale,
$\alpha^{\prime -1/2}$, we will establish several new results in
this paper. First, we show that in either ensemble the growth of the free energy, 
$F(\beta)$, at high temperatures far beyond the string
scale is only as fast as in a 2d quantum field theory. Thus, as
was conjectured with only limited intuition as far back as 1988 by
Atick and Witten for closed string theories \cite{aw}, there is a dramatic reduction in the
{\em growth} of the free energy at high temperatures. 
More recently \cite{polchinskibook}, Polchinski has shown that the
$T^2$ growth in the free energy at high temperatures is
a direct consequence of the thermal self-duality of the vacuum
functional in the closed bosonic string theory. We will show in this 
paper that Polchinski's observation
extends to both the heterotic closed, and type I unoriented open
and closed string ensembles. In either case, the $T^2$ 
growth in the string free energy
at high temperatures follows as a direct consequence of
the Euclidean T-duality transformations that link the
thermal ground states of the supersymmetric string theories in
pairs. Furthermore, we will find that the full high temperature 
expansion in powers of $\beta$ for the string free energy can be obtained
explicitly upon term-by-term evaluation of the
one-loop modular integrals in the string mass level expansion.

\vskip 0.1in Our starting point in this paper is the generating
functional of connected one-loop vacuum string graphs, $W(\beta)$
$\equiv$ ${\rm ln}$ ${\rm Z}(\beta)$, where ${\rm Z}(\beta)$ is
the canonical partition function. $W(\beta)$ is derived from first
principles in the Polyakov path integral formalism following
\cite{poltorus}.The free energy, $F(\beta)$, and
vacuum energy density, $\rho(\beta)$, can be directly inferred from 
$W(\beta)$. Let us
recall the basic thermodynamic identities of the canonical
ensemble \cite{pippard}:
\begin{equation}
F= -W/\beta = V \rho , \quad P =
      - \left ( {{\partial F}\over{\partial V}} \right )_T , \quad
      U = T^2 \left ( {{\partial W}\over{\partial T}} \right )_V
, \quad
      S = - \left ( {{\partial F}\over{\partial T}} \right )_V ,
 \quad
      C_V = T \left ( {{\partial S}\over{\partial T}} \right )_V
\quad . \label{eq:freene}
\end{equation}
Note that $W(\beta)$ is an intensive thermodynamic variable
without explicit dependence on the spatial volume. $F$ is the
Helmholtz free energy of the ensemble of strings, $U$ is the
internal energy, and $\rho$ is the finite temperature effective
potential, or vacuum energy density, at finite temperature. $S$ and
$C_V$ are, respectively, the entropy and specific heat of the
canonical ensemble. The pressure of the string ensemble simply
equals the negative of the vacuum energy density, as is true for a
cosmological constant, just as in an ideal fluid with negative
pressure \cite{pippard}. The enthalpy, $H$$=$$U$$+$$PV$,
the Helmholtz free energy, $F$$=$$U$$-TS$, and the Gibbs function,
also known as the Gibbs free energy, is $G$$=$$U$$-$$TS$$+$$PV$.
As a result of these relations, {\em all} of the thermodynamic
potentials of the string ensemble have been give a simple,
first-principles, formulation in terms of the Polyakov path integral. 
Notice, in particular, that since $P$$=$$-\rho$, the
one-loop contribution to the Gibbs free energy of the string
ensemble vanishes identically! 

\vskip 0.1in We should address the expected Jeans
instability of a gravitating statistical ensemble \cite{pippard,gyp,aw}, and
the basic definition of the thermodynamic limit in the presence of 
gravity.  Consider an ensemble of total mass, $M$,  and
Schwarschild radius, $R_S$, with one-loop vacuum energy 
density, $\rho$ $=$ $F(\beta)/R^{D-1}$. Recall that the Newtonian
gravitational coupling, $G_N$$\simeq$$g_s^2$, where $g_s$ is the
closed string coupling. Since $M$ $\sim$ $\rho R_S^{D-1}$,
we have the relation:
\begin{equation}
R_S >> \left ( {{1}\over{g_s^2 \rho (\beta) }} \right )^{1/2} \quad .
\label{eq:schw}
\end{equation}
In other words, strictly speaking, it is only possible to take the infinite volume limit
$V$ $\to$ $\infty$, $\alpha^{\prime}$ $\to$ $0$, when the energy density per unit 
volume of the gravitating ensemble happens
to be zero, or in the limit of weak string coupling \cite{gyp,aw,garyp}. 

\section{The Impossibility of a Type II Canonical Ensemble}

\vskip 0.1in We will begin by explaining why there can be no equilibrium type II
superstring ensemble in the absence of Dbranes or background fluxes, either
of which introduces Yang-Mills gauge fields in the low energy field theory limit 
of the finite temperature superstring theory. 

\vskip 0.1in As explained at the outset in this paper, an equilibrium ensemble of
type II strings requires that we identify a tachyon-free nonsupersymmetric
background at all temperatures starting
from zero. Let us examine some consequences of finding such a solution.
Our interest is in a one-parameter family of nonsupersymmetric backgrounds of
the type II superstring with Euclidean target spacetime $R^9$$\times$$S^1$; 
a line of fixed points of the worldsheet RG 
parametrized by the single parameter, $T$, the inverse of the circumference
of the $S^1$, whose physical interpretation is temperature. 
Continuity of the vacuum functional as a function of the parameter 
$T$ requires
that, at least for small values of $T$, $W(T)$ 
take the form of an integral over the
modular group of genus one Riemann surfaces with a modular
invariant integrand. In addition, the leading terms in
the closed string mass level expansion for small $T$ 
must have a self-consistent interpretation in terms of the field theoretic
modes of a {\em 10D supergravity field theory at finite
temperature}. Thus, the target space supersymmetry of the zero temperature
vacuum must be
spontaneously broken at all temperatures different from zero. And the existence of a
stable gravitating ensemble in thermodynamic equilibrium requires
the absence of tachyonic thermal modes in the string mass spectrum.
Finally,  we must check that the vacuum energy density satistfies the criterion for
the absence of the classical Jeans instability in a gravitating
ensemble \cite{gyp,aw}. Thus, we can succinctly state the
infrared consistency conditions required of an equilibrium ensemble of
type II strings at finite temperatures:
\begin{itemize}
 \item the absence of thermal tachyons in the mass level expansion.
 \item the spontaneous breaking of target space supersymmetry at
 low temperatures, $T$ $>$ $0$.
 \item the demonstration of a $T^{10}$ growth of the free energy at 
 low temperatures, when we
 isolate the leading 
 contribution to the string mass level expansion 
 from the low energy, field theoretic supergravity modes alone.
 \end{itemize}
We will find that the first two conditions are incompatible: modular
invariance of the one-loop type II vacuum amplitude turns out to be 
extremely restrictive. The spontaneous breaking of supersymmetry in the 
10D type II superstring vacuum at temperatures different from zero necessarily implies
the existence of a whole slew of {\em low temperature tachyonic momentum 
modes}. This invalidates the possibility of a type II string canonical ensemble,
even at low temperatures far below the string scale.

\vskip 0.1in In order to understand the stringent limitations on
the expression for the one-loop vacuum amplitude in the finite
temperature vacuum imposed by modular
invariance, recall that the type II string mass level expansion results from a
generic combination of the
four, holomorphic and anti-holomorphic, Jacobi theta functions,
weighted by, a priori, undetermined phases. In addition, the zero
temperature mass formulae in each spin structure sector will now
acquire additional contributions from thermal winding and thermal
momentum modes. Let us begin by examining the shifts in
the ground state energy of the NS-NS vacuum of the type II superstring
at finite temperature. Recall that this state is
tachyonic in the zero temperature Hilbert space, eliminated
from the mass level expansion for either supersymmetric type II
superstring theory by a judicious choice of phases in the vacuum
amplitude. It will be easy to verify that there is no corresponding 
choice of modular invariant 
phases that can eliminate the tachyonic thermal modes of the 
NS-NS sector, while spontaneously breaking supersymmetry at all
temperatures different from zero.

\vskip 0.1in Suppressing the oscillator contributions, the ground
state energy in the NS-NS sector for physical states satisfying
the level matching constraint takes the form:
\begin{equation}
({\rm mass})_L^2 = ({\rm mass})_R^2 = {{4}\over{\alpha^{\prime}}}
 \left [ - \half + \half \left (
{{4 \pi^2 \alpha^{\prime} n^2 }\over{\beta^2}} + {{\beta^2
w^2}\over{4\pi^2 \alpha^{\prime} }}  \right ) \right ]  \quad .
\label{eq:massIj}
\end{equation}
Notice that the pure momentum and pure winding states, $(n,0)$ and
$(0,w)$, are potential tachyons that enter into the general
expression for the level expansion. In the absence of oscillator
excitations, each pure momentum mode, $(\pm n, 0)$, is tachyonic
{\em upto} some critical temperature, $T^2_n$ $=$ $1/2 n^2\pi^2
\alpha^{\prime} $, beyond which it turns marginal (massless).
Conversely, each pure winding mode $(0,\pm w)$, turns tachyonic
{\em beyond} some critical temperature, $T^2_w$ $=$ $w^2/8\pi^2
\alpha^{\prime}$. Finally, recall that under a thermal duality
transformation, the type IIA superstring is mapped to the type IIB
superstring:
\begin{equation}
\beta_{\rm IIA} \to \beta_{\rm IIB} = 4 \pi^2 \alpha^{\prime}
/\beta_{\rm IIA} , \quad \quad (n,w)_{\rm IIA} \to
(n'=w,w'=n)_{\rm IIB} \quad , \label{eq:dual}
\end{equation}
thus interchanging the identification of momentum and winding
modes in the type IIA and type IIB thermal specta. In other words,
in the absence of a 
nontrivial Ramond-Ramond sector, 
the result for the IIA and IIB one-loop vacuum functionals 
always coincides. In either case, 
upon compactification on the
circle of radius $\beta/2\pi$, the zero mode spectrum takes the form
\cite{poltorus, polchinskibook}:
\begin{equation}
p_L = {{2\pi n}\over{\beta}} + {{w \beta}\over{2 \pi
\alpha^{\prime} }} , \quad p_R = {{2\pi n}\over{\beta}} - {{w
\beta}\over{2 \pi \alpha^{\prime} }}  \quad , \label{eq:mom}
\end{equation}
so that the contribution to the path integral from the $(n,w)$th 
sector is:
\begin{equation}
{\rm exp} \left [ - \pi \tau_2 \left ( {{4\pi^2 \alpha' n^2
    }\over{\beta^2}}    + {{ w^2 \beta^2}\over{4\pi^2 \alpha'}} \right )  + 2 \pi i n w \tau_1 \right ] 
    \quad .
\label{eq:modethr}
\end{equation}
Thus, upon compactifying either type II superstring on $R^9$$\times$$S^1$,
the one-loop vacuum
amplitude takes the form:
\begin{eqnarray}
W_{\rm II} (\beta) &&=  \beta L^{9} (4\pi^2 \alpha^{\prime})^{-5}  \int_{\cal F}
{{d^2 \tau}\over{4\tau_2^2}} \cdot 
  (\tau_2)^{-8/2} [\eta(\tau) {\bar{\eta}} ({\bar{\tau}} )]^{-8/2} \cr
   &&\quad \times     {{1}\over{4}} 
  \left [  ({{\Theta_{00}}\over{\eta}})^4 - ({{ \Theta_{01}}\over{\eta}})^4 - ({{\Theta_{10}}\over{\eta}})^4 \pm
  ({{\Theta_{11}}\over{\eta}}) ^4 \right ] \left [
   ( {{ {\bar{\Theta}}_{00} }\over { {\bar{\eta}} }})^4 - ({{{\bar{\Theta}}_{01} }\over { {\bar{\eta}} }})^4  - ({{ {\bar{\Theta}}_{10} }\over{ {\bar{\eta}} }})^4
   \pm  ({{ {\bar{\Theta}}_{11} }\over{ {\bar{\eta}} }})^4 
\right ] \cr
&& \quad \quad \quad \times \sum_{n,w=-\infty}^{\infty} {\rm exp} \left [ - \pi \tau_2 \left ( {{4\pi^2 \alpha' n^2
    }\over{\beta^2}}    + {{ w^2 \beta^2}\over{4\pi^2 \alpha'}} \right )  + 2 \pi i n w \tau_1 \right ] 
 \quad ,
\label{eq:IIs}
\end{eqnarray}
Notice that the familiar choice of phase that projects out
the tachyonic NS-NS vacuum at zero temperature, namely, the relative minus 
sign between (00) and (01) spin structure sectors,  
also suffices to eliminate all of the tachyonic
momentum and winding modes from contributing to the string mass level expansion.
This is the precise type II analog of the thermal tachyon spectrum of the closed bosonic string 
ensemble \cite{poltorus,polchinskibook}, except that they have been rendered
{\em unphysical} due to the choice of phase. In addition, we find that modular invariance 
has simultaneously forced the particular combination 
of worldsheet fermionic spin structures displayed
in the expression above. Thus, target spacetime supersymmetry is unbroken!
In other words, there is no viable type II string canonical ensemble. We need to 
introduce a new feature into the one-loop string vacuum amplitude that can loosen up the 
stringent constraints on the combination of fermionic spin structures arising
from
modular invariance. This new feature turns out to be the introduction of Yang-Mills gauge 
fields and, consequently, the possibility of spontaneous supersymmetry breaking via a
temperature dependent Wilson line \cite{decon,fermionic,holo}.

\section{Free Energy of the Heterotic Canonical Ensemble}

\vskip 0.1in Unlike the type II superstrings where the constraints from
modular invariance proved far too restrictive to permit a tachyon-free
thermal ground state, the heterotic string theory has a Yang-Mills sector.
This introduces the possibility of a Wilson line gauge background, loosening
the constraints from modular invariance while enabling 
a tachyon-free thermal vacuum and, consequently, a self-consistent
formulation of the equilibrium canonical ensemble. Consider the
ten-dimensional supersymmetric $E_8$$\times$$E_8$ theory at zero
temperature. The $\alpha'$$\to$$0$ low energy field theory limit
is 10D $N$$=$$1$ supergravity coupled to $E_8$$\times$$E_8$
Yang-Mills gauge fields. What happens to the supersymmetric ground
state of this theory at finite temperature? 
We wish to derive an analogue of the zero temperature vacuum functional 
which describes 
a stable finite temperature ground state:  thermal 
tachyons should be absent, target space supersymmetry must be spontaneously broken
at finite temperature, and the free energy must grow as $T^{10}$ at
low temperatures when only massless field theory modes are excited.
Most
importantly, since we wish to preserve the finiteness and perturbative
renormalizability of the
zero temperature ground state
at finite temperature, it is
important to preserve the invariance of the 
one-loop 
vacuum functional under the modular group of the torus. 
Finally, we must 
require self-consistency with the thermal duality transformations: 
the $E_8$$\times$$E_8$ and $\rm Spin(32)/Z_2$ heterotic
string theories are related by the Euclidean timelike T-duality
transformation: 
\begin{equation}
\beta_{\rm E_8 \times E_8 } \to \beta_{\rm SO(32)} = 4 \pi^2 \alpha^{\prime}
/\beta_{\rm E_8 \times E_8 } , \quad \quad (n,w)_{\rm E_8 \times E_8 } \to
(n'=w,w'=n)_{\rm SO(32) } \quad , \label{eq:duals}
\end{equation}
thus interchanging the identification of momentum and winding
modes in the  $E_8$$\times$$E_8$ and $SO(32)$ thermal spectra. 
Thus, the finite temperature vacuum functional is
required to interpolate between the following two spacetime
supersymmetric limits: in the $\beta$$\to$$\infty$ limit we
recover the vacuum functional of the supersymmetric
$E_8$$\times$$E_8$ heterotic string, while in the $\beta$$=$$0$
limit we must recover, instead, the vacuum functional of the
supersymmetric $\rm Spin(32)/Z_2$ heterotic string. 

\vskip 0.1in  
It is helpful to begin by examining the resulting shifts in
the ground state energy of the (NS$+$,NS$+$) vacuum due to
the winding and momentum modes in the mass 
spectrum obtained upon compactification on the circle of radius $\beta/2\pi$. 
Suppressing the oscillator contributions, the ground
state energy in this sector for physical states satisfying
the level matching constraint takes the form:
\begin{equation}
({\rm mass})_L^2 = ({\rm mass})_R^2 = {{4}\over{\alpha^{\prime}}}
 \left [ - 1 + \half {\bf k}_L^2  + \half \left (
{{4 \pi^2 \alpha^{\prime} n^2 }\over{\beta^2}} + {{\beta^2
w^2}\over{4\pi^2 \alpha^{\prime} }}  \right ) \right ] = 
{{4}\over{\alpha^{\prime}}}  \left [ - \half  + \half \left (
{{4 \pi^2 \alpha^{\prime} n^2 }\over{\beta^2}} + {{\beta^2
w^2}\over{4\pi^2 \alpha^{\prime} }}  \right ) \right ]  ,
\label{eq:massII}
\end{equation}
where ${\bf k}_L$ belongs in the lattice $E_8$$\times$$E_8^{\prime}$, and either
$E_8$ lattice is spanned by the vectors:
\begin{eqnarray}
(n_1 , \cdots , n_8), ~ (n_1 + \half , \cdots , n_8 + \half ), \quad 
 {\rm where ~ all ~ n_i \in  Z, ~ and ~ \sum_{i=1}^8 n_i \in 2Z} \quad .
\label{eq:latthe8}
\end{eqnarray}
The corresponding result for the ${\rm Spin(32)/Z_2}$ lattice takes the
form:
\begin{eqnarray}
(n_1 , \cdots , n_{16} ), ~ (n_1 + \half , \cdots , n_{16} + \half ), \quad 
 {\rm where ~ all ~ n_i \in  Z, ~ and ~ \sum_{i=1}^{16} n_i \in 2Z} \quad .
\label{eq:latthe16}
\end{eqnarray}
As for the type II superstrings, the pure momentum and pure winding states, $(n,0)$ and
$(0,w)$, are potential tachyons that enter into the general
expression for the level expansion. In the absence of oscillator
excitations, each pure momentum mode, $(\pm n, 0)$, is tachyonic
{\em upto} some critical temperature, $T^2_n$ $=$ $1/2 n^2\pi^2
\alpha^{\prime} $, beyond which it turns marginal (massless).
Conversely, each pure winding mode $(0,\pm w)$, turns tachyonic
{\em beyond} some critical temperature, $T^2_w$ $=$ $w^2/8\pi^2
\alpha^{\prime}$. Notice that retaining the relative minus sign between the
contributions from (00) and (01) sectors of the right-moving
superconformal field theory in the supersymmetric vacuum 
functional, as in the previous section, would eliminate both $O((q{\bar{q}})^{1/2})$
contributions to the level expansion, in addition to all of the $\beta$ dependent
tachyonic thermal modes. Thus, for example, {\em all} of the potential tachyons in the
(NS$+$,NS$+$) sector of the circle compactified supersymmetric ground state 
are eliminated from the level expansion for physical states
in one fell swoop. But can this mechanism work in the absence of
spacetime supersymmetry? We will need to identify a different chiral 
modular invariant, one that spontaneously breaks spacetime supersymmetry
at finite temperature. 
The key lies in the introduction of a 
{\em temperature dependent } Wilson line background \cite{decon,fermionic,holo}.

\subsection{Axial Gauge and the Euclidean Timelike Wilson Line}

\vskip 0.1in  The generating functional of connected one-loop
vacuum string graphs in the stable finite temperature vacuum will be
given by an expression of the form:
\begin{equation}
W_{\rm het} (\beta)= \beta L^9 (4\pi^2 \alpha^{\prime})^{-5} \int_{\cal F}
{{d^2 \tau}\over{4\tau_2^2}}
  \tau_2^{-4}   Z_{\rm het} (\beta)
\quad , \label{eq:het}
\end{equation}
where the function $Z_{\rm het}(\beta)$
denotes the desired level expansion for
the thermal mass spectrum of
the heterotic ensemble at generic temperatures that can satisfy the
interpolations to the zero temperature limit required by the thermal
duality transformations. In particular, we wish to
identify a suitable interpolating expression for $W_{\rm het}(\beta)$
which satisfies the infrared consistency conditions
 at generic values of $\beta$, matching smoothly with the
known vacuum functional of the supersymmetric $E_8$$\times$$E_8$
string theory at zero temperature ($\beta$$=$$\infty$):
\begin{eqnarray}
 W (\beta)|_{\rm T=0} =&&
     \beta L^{9} (4\pi^2 \alpha^{\prime})^{-5} \int_{\cal F}
\left \{ {{d^2 \tau}\over{4\tau_2^2}} \cdot 
  (\tau_2)^{-4}  [ \eta(\tau) {\bar{\eta}} ({\bar{\tau}} )]^{-8} \right \}
  \cr &&\quad \quad \times 
  \left [   \{  ({{\Theta_{00}}\over{\eta}})^4 - ({{ \Theta_{01}}\over{\eta}})^4 \} - 
   \{ ({{\Theta_{10}}\over{\eta}})^4 \pm
  ({{\Theta_{11}}\over{\eta}}) ^4 \} \right ]  \cr
  && \quad \quad \quad \quad \times
  {{1}\over{4}} \left [  \left ({ { \Theta_{00} }\over{ \eta }} \right )^{8}
 + \left ({ { \Theta_{01} }\over{\eta }} \right )^{8} + 
\left ({ { \Theta_{10} }\over{ \eta }} \right )^{8} + \left ({ { \Theta_{11} }\over{ \eta }} \right )^{8}
 \right ]^2 \quad , 
\label{eq:e8e8}
\end{eqnarray}
while also recovering the vacuum functional of the T-dual supersymmetric
${\rm Spin } (32)/{\rm Z}_2$
heterotic string {\em in the presence of the T-dual Wilson line background}:
\begin{eqnarray}
 W (\beta)|_{\rm T^\prime= 0} =&&
     \beta L^{9} (4\pi^2 \alpha^{\prime})^{-5} \int_{\cal F}
\left \{ {{d^2 \tau}\over{4\tau_2^2}} \cdot 
  (\tau_2)^{-4}   [ \eta(\tau) {\bar{\eta}} ({\bar{\tau}} )]^{-8} \right \}
  \cr &&\quad \quad \times 
  \left [   \{  ({{\Theta_{00}}\over{\eta}})^4 - ({{ \Theta_{01}}\over{\eta}})^4 \} - 
   \{ ({{\Theta_{10}}\over{\eta}})^4 \pm
  ({{\Theta_{11}}\over{\eta}}) ^4 \} \right ]  \cr
  && \quad \quad \quad \quad \times
  {{1}\over{4}} \left [  \left ({ { \Theta_{00} }\over{ \eta }} \right )^{16}
 + \left ({ { \Theta_{01} }\over{\eta }} \right )^{16} + 
\left ({ { \Theta_{10} }\over{ \eta }} \right )^{16} + \left ({ { \Theta_{11} }\over{ \eta }} \right )^{16}
 \right ] \quad .
\label{eq:spin32}
\end{eqnarray}
It turns out that the desired expression 
for ${\rm Z}_{\rm het}(\beta)$, and the 
pair of T-dual Wilson line backgrounds necessary for
this interpolation, 
can be inferred using extant results 
in the heterotic string literature. 

\vskip 0.1in The argument proceeds as follows. The modular invariant
possibilities for the sum over spin structures in the 10d
heterotic string have been classified, both by free fermion and by
orbifold techniques \cite{sw,dh,agmv,klt}, and there is a {\em
unique} nonsupersymmetric and tachyon-free ground state 
with gauge
symmetry $SO(16)$$\times$$SO(16)$. In the fermionic
formulation, the GSO projection
differs from that of the 
$E_8$$\times$$E_8$ heterotic string as follows:
\begin{equation}
e^{\pi {\tilde{F}} + \alpha + \alpha^{\prime} } = e^{\pi i F + \alpha^{\prime} +{\tilde{\alpha}} } = 
e^{\pi i {\tilde{F'}} + {\tilde{\alpha}} + \alpha }= +1 \quad ,
\label{eq:torsion}
\end{equation}
where $\alpha$, $\alpha^\prime$, and ${\tilde{\alpha}}$, equal,
$+1$, or $0$, respectively, in the Ramond, or Neveu-Schwarz, sector
for each of the three 8-fermion blocks: right-moving, left-moving 
gauge, and left-moving
gauge$^\prime$
\cite{agmv,polchinskibook}.
The \lq\lq all-positive" choice of worldsheet chiralities in every sector of the 
supersymmetric $E_8$$\times$$E_8$ theory are
therefore altered to the more complicated, and nonsupersymmetric, projection:
\begin{eqnarray}
&&(\rm NS+, NS+, NS+), (\rm NS- , NS- , R+ ), (\rm NS-, R+, NS-), (\rm R+, NS- , NS- ),
\cr
&&(\rm NS+, R-, R-), (\rm R- , NS+ , R- ), (\rm R-, R-, NS+), (\rm R+ , R+ , R+ ) \quad .
\label{eq:chiralities}
\end{eqnarray}
This choice of twisted boundary conditions introduces discrete torsion in the, 
alternative, orbifold description of this theory \cite{dh,polchinskibook}, raising 
the ground state energy in the sectors that previously provided the massless 
chiral fermions of the supersymmetric heterotic string. Most illuminating, perhaps,
is to understand this theory as the result of taking the noncompact limit of 
either circle-compactified $E_8$$\times$$E_8$ or ${\rm Spin(32)/Z_2}$ heterotic string
theories, upon 
introduction of a Wilson line background 
that spontaneously breaks both spacetime 
supersymmetry, as well as breaking the nonabelian gauge
symmetry to $SO(16)$$\times$$SO(16)$. This interpretation is clarified by invoking
the interpolating vacuum functional approach \cite{bgsrohm,itoyama,gv,ginine}.

\vskip 0.1in
Begin with the one-loop vacuum functional of the
circle compactified nonsupersymmetric and 
tachyon free $SO(16)$$\times$$SO(16)$ heterotic string theory \cite{agmv},
 where the $S^1$ is
a circle of radius $\beta/2\pi$:
\begin{eqnarray}
 W_{\rm SO(16) \times SO(16) } (\beta) =&&
      \beta L^{9} (4\pi^2 \alpha^{\prime})^{-5} \int_{\cal F}
\left \{ {{d^2 \tau}\over{4\tau_2^2}} \cdot 
  (\tau_2)^{-4}   [ \eta(\tau) {\bar{\eta}} ({\bar{\tau}} )]^{-8} \right \}
  \cr &&\quad \quad \times 
   {{1}\over{4}} \left [
({{\Theta_2}\over{\eta}})^8 ({{\Theta_4}\over{\eta}})^8
({{{\bar{\Theta_3}}}\over{\eta}})^4 - ({{\Theta_2}\over{\eta}})^8
({{\Theta_3}\over{\eta}})^8 ({{{\bar{\Theta_4}}}\over{\eta}})^4 -
({{\Theta_3}\over{\eta}})^8 ({{\Theta_4}\over{\eta}})^8
({{{\bar{\Theta_2}}}\over{\eta}})^4 \right ]
     \cr
&& \quad \quad \quad \quad \quad \quad \quad \times ~  \sum_{n,w=-\infty}^{\infty}  {\rm exp} \left [ - \pi \tau_2 \left ( {{4\pi^2 \alpha' n^2
    }\over{\beta^2}} + {{w^2 \beta^2}\over{4\pi^2 \alpha'}} \right )  + 2 \pi i n w \tau_1 \right ]
 \quad .
\label{eq:so16}
\end{eqnarray}
It is easy to verify by inspection that the integrand in this expression is invariant under 
the one-loop modular group. The expression also satisfies some of 
the infrared consistency conditions
for a viable thermal ground state outlined in section 2: there are no thermal tachyons. 
However, the noncompact $\beta $ $\to$ $0$ limit 
of this expression does not lead to a spontaneous 
restoration of target spacetime supersymmetry since we recover, 
instead,  the nonsupersymmetric $O(16)$$\times$$O(16)$ 
10D modular invariant. We must incorporated a temperature dependent
Wilson line background in order to interpolate between 10D supersymmetric
and 9D nonsupersymmetric ground states.

\vskip 0.1in
The spontaneous restoration of spacetime supersymmetry in the 
noncompact limit within the larger family of one-parameter continuously 
connected 9D $SO
(16)$$\times$$SO(16) $
heterotic string vacua {\em with nontrivial Wilson
line gauge backgrounds} was discovered in independent
numerical
investigations by Itoyama and Taylor \cite{itoyama}
and Ginsparg and Vafa  \cite{nsw,ginine,gv}. The plots in the
latter paper indicate clearly the distinction between taking the noncompact
limit of the expression in Eq.\ (\ref{eq:so16}): where the vacuum energy 
density goes thru a minimum at the self-dual point without touching zero,
and taking the noncompact
limit in the presence of a Wilson line background. It is in the latter case that the  
target spacetime supersymmetry is spontaneously restored at $T$ $=$ $0$.
Additional massless gauge bosons appear 
as we approach the $T$ $=$ $0$ limit, such that we recover the full 
$SO(32)$ gauge symmetry \cite{gv}. 
Recall that the 
gauge lattice vector  
${\bf k}_L$ in the equivalent worldsheet 
bosonic description of $W_{\rm het} (\beta)$
will belong in a $(1,17)$-component self-dual lattice of
Lorentzian signature.
Generic points in the lattice can be reached by Lorentz boosts that preserve the 
Lorentzian self-duality property, a pre-requisite for modular invariance, and each 
describes a heterotic string vacuum that can be
reached by continuous interpolation from the theory above.
The precise form of a generic vector in the self-dual lattice $\Gamma^{(1,17)}$ is given
by \cite{nsw,ginine,polchinskibook}: 
\begin{equation}
(p^0_R ,  p^0_L , {\bf k}_L )  \to  (p^{0 \prime}_R , p^{0\prime}_L , {\bf k}_L^{\prime} ) =
(p_R^0 - {\bf k}\cdot {\bf A}^0
- {{w^0 \beta}\over{4\pi}} {\bf A}_0 \cdot {\bf A}^0 ,  p^0_L - {\bf k}\cdot {\bf A}^0 -
{{w^0 \beta}\over{4\pi}} {\bf A}_0 \cdot {\bf A}^0 ,  {\bf k}_L + w^0 \left ( {{\beta}\over{2\pi}} \right ) {\bf A}_0 ) \quad , \label{eq:boost}
\end{equation}
where:
\begin{equation}
p^0_L = {{2\pi n}\over{\beta}} + {{w \beta}\over{2 \pi
\alpha^{\prime} }} , \quad p^0_R = {{2\pi n}\over{\beta}} - {{w
\beta}\over{2 \pi \alpha^{\prime} }}  , \quad \quad n,w \in {\rm Z^+} \quad .
\label{eq:scalemom}
\end{equation}

\vskip 0.1in
In particular, it is illuminating to consider the two lattice boosts that, respectively, interpolate
between the nonsupersymmetric and tachyon-free 9D $SO(16)$$\times$$SO(16)$ 
vacuum functional and the pair of 10D $E_8$$\times$$E_8$, and $SO(32)$,
supersymmetric string limits. The result for the required lattice-boosts is due to Ginsparg \cite{ginine}.
Starting with the supersymmetric 10D ${\rm Spin(32)/Z_2 }$ string, compactify on
a circle of radius $\beta/2\pi$ with Wilson line gauge background:
$\beta {\bf A}^0 $ $=$ $ 2\pi (1^8 ;  0^8 )  $.
The closed form expression
for the interpolating one-loop vacuum functional describing the 
exactly marginal flow to the 10D ${\rm Spin}(32)/{\rm Z}_2$ heterotic
string vacuum
can be found in the paper of
Itoyama and Taylor  \cite{itoyama}\footnote{The vacuum functional of interest to us has been
called the \lq\lq Twist II" model by the authors of \cite{itoyama}, as is clear
from the plot in Figure 1 of their paper. 
The T-duality transform of the expression in Eq.\ (13) 
of Ref.\ [\cite{itoyama}] gives the desired expression for ${\rm Z}_{\rm het} (\beta)$; a typo
in the last line of Eq.\ (13) has been corrected. The flow
to the 10D $E_8$$\times$$E_8$ vacuum, instead, would be described by the interpolation with 
the T-dual Wilson line background, as in \cite{ginine}.}:
\begin{eqnarray}
 Z_{\rm het } (\beta) =&& {{1}\over{8}}  [ \eta(\tau) {\bar{\eta}} ({\bar{\tau}} )]^{-8} [\eta(\tau)]^{-16} 
 [{\bar{\eta}} ({\bar{\tau}})]^{-4}  \times 
  \{  \left ( {\cal E}_0 + {\cal E}_{1/2} \right ) \left [
 {\bar{\Theta}}_3^4 \Theta_4^8 \Theta_2^8 - {\bar{\Theta}}_4^4 \Theta_2^8
\Theta_3^8  -
{\bar{\Theta}}_2^4 \Theta_3^8 \Theta_4^8  \right ] 
\cr && \cr && \quad
 + \left ( {\cal O}_0 - {\cal O}_{1/2} \right )  \left [ \half {\bar{\Theta}}_3^4 
\left ( 
\Theta_2^{16} +  \Theta_3^{16} + \Theta_4^{16}  \right ) 
- {\bar{\Theta}}_4^4 
\Theta_3^{8}  \Theta_4^{8} 
- {\bar{\Theta}}_2^4 
\Theta_2^{8} \Theta_3^{8} \right ]
     \cr &&\cr 
&& \quad  \quad\quad + 
\left ( {\cal E}_0 - {\cal E}_{1/2} \right )  \left [  - \half {\bar{\Theta}}_2^4 
\left ( 
\Theta_2^{16} +  \Theta_3^{16} + \Theta_4^{16}  \right ) 
- {\bar{\Theta}}_4^4 
\Theta_4^{8} \Theta_2^{8}  +  {\bar{\Theta}}_3^4 
\Theta_2^{8}  \Theta_3^{8}  \right ] 
\cr
&& \cr && \quad \quad \quad \quad \quad   + 
\left ( {\cal O}_0 + {\cal O}_{1/2} \right )  \left [- \half {\bar{\Theta}}_4^4 
\left ( 
\Theta_2^{16} +  \Theta_3^{16} + \Theta_4^{16}  \right ) 
+ {\bar{\Theta}}_3^4 
\Theta_3^{8}  \Theta_4^{8} 
- {\bar{\Theta}}_2^4 
\Theta_2^{8} \Theta_4^{8} \right ] \} 
 \quad  . \cr &&
\label{eq:so1twi}
\end{eqnarray}
As explained above, we use the T-dual of the
standard basis for thermal mode summations introduced in \cite{bgsrohm}, 
in terms of which the one-loop modular transformations are especially easy to 
verify:
\begin{equation}
{\cal E}_0 :  n  \in {\rm Z} ,  ~ w ~ {\rm even} . ~ 
{\cal E}_{1/2} :   n  \in {\rm  Z + 1/2 },  ~ w ~ {\rm even} . ~ 
{\cal O}_0 :   n  \in {\rm Z} ,  ~ w ~ {\rm odd} . ~ 
{\cal O}_{1/2} :  n  \in {\rm  Z + 1/2} ,  ~ w ~ {\rm odd} .
\label{eq:modn}
\end{equation}
A $\tau$$\to$$-1/\tau$ transformation leaves $({\cal E}_0$$+$${\cal E}_{1/2})$ and
$({\cal O}_0$$-$${\cal O}_{1/2})$ invariant, while interchanging $({\cal E}_0$$-$${\cal E}_{1/2})$
and $({\cal O}_0$$+$${\cal O}_{1/2})$. Under $\tau$$\to$$\tau$$+$$1$, the first three 
summations are invariant, while ${\cal O}_{1/2}$ maps to the negative of itself.
It is easy to verify the invariance of this expression under a $\tau$ $\to$ $-1/\tau$
transformation. Under a $\tau$ $\to$ $\tau +1$ transformation, the first and third lines
within curly brackets transform with a minus sign, while the second and fourth are 
interchanged with a minus sign. The overall minus sign is accounted for by the transformation
of the eta functions. As before,
$w$ and $n$ denote thermal windings, and thermal momenta, respectively. Notice 
the appearance of
both half-integer and integer Matsubara frequencies along this flow. 
Notice, also, that because
of the inclusion of a Wilson line background, the interpolating functional describing the flow to
the T-dual $E_8$$\times$$E_8$ vacuum cannot be inferred directly from the expression 
above, because the T-duality acts nontrivially on the gauge field, in addition to interchanging
momenta and windings. We must start afresh with the
10D $E_8$$\times$$E_8$ vacuum 
functional, compactifying on a circle of radius $\beta$
with temperature dependent 
Wilson line background as before. The interpolation takes the form:
\begin{eqnarray}
 Z_{\rm het } (\beta) =&& {{1}\over{8}}  [ \eta(\tau) {\bar{\eta}} ({\bar{\tau}} )]^{-8} [\eta(\tau)]^{-16} 
 [{\bar{\eta}} ({\bar{\tau}})]^{-4}  \times 
  \{  \left ( {\cal E}_0 + {\cal E}_{1/2} \right ) \left [
 {\bar{\Theta}}_3^4 \Theta_4^8 \Theta_2^8 - {\bar{\Theta}}_4^4 \Theta_2^8
\Theta_3^8  -
{\bar{\Theta}}_2^4 \Theta_3^8 \Theta_4^8  \right ] 
\cr && \cr && \quad
 + \left ( {\cal O}_0 - {\cal O}_{1/2} \right )  \left [ \half {\bar{\Theta}}_3^4 
\left ( 
\Theta_2^{8} +  \Theta_3^{8} + \Theta_4^{8}  \right )^2 
- {\bar{\Theta}}_4^4 
\Theta_3^{8}  \Theta_4^{8} 
- {\bar{\Theta}}_2^4 
\Theta_2^{8} \Theta_3^{8} 
\right ]
     \cr &&\cr 
&& \quad  \quad\quad + 
\left ( {\cal E}_0 - {\cal E}_{1/2} \right )  \left [  - \half {\bar{\Theta}}_2^4 
\left ( 
\Theta_2^{8} +  \Theta_3^{8} + \Theta_4^{8}  \right )^2 
- {\bar{\Theta}}_4^4 
\Theta_4^{8} \Theta_2^{8}  +  {\bar{\Theta}}_3^4 
\Theta_2^{8}  \Theta_3^{8}  \right ] 
\cr
&& \cr && \quad \quad \quad \quad \quad   + 
\left ( {\cal O}_0 + {\cal O}_{1/2} \right )  \left [- \half {\bar{\Theta}}_4^4 
\left ( 
\Theta_2^{8} +  \Theta_3^{8} + \Theta_4^{8}  \right )^2 
+ {\bar{\Theta}}_3^4 
\Theta_3^{8}  \Theta_4^{8} 
- {\bar{\Theta}}_2^4 
\Theta_2^{8} \Theta_4^{8} \right ] \} 
 \quad  . \cr &&
\label{eq:so1tw}
\end{eqnarray}

\vskip 0.1in It should be noted that in the presence of the timelike Wilson line backgrounds, 
the Yang-Mills gauge group is no longer a full $SO(32)$, nor a full $E_8$$\times$$E_8$, 
except in their respective noncompact limits: $\beta$ $\to$ $0$.  At generic values of  $\beta$, 
we have the gauge group $SO(16)$$\times$$SO(16)$$\times$$U(1)$. Enhanced 
gauge symmetries of rank $17$ will appear
as we vary $(\beta, A_0(\beta))$, subject to the Lorentzian (17,1) self-duality conditions
that ensure modular invariance. This is
the usual phenomenon of Wilson line gauge symmetry breaking, a hallmark of heterotic
string phenomenology. From the perspective of the low energy Yang-Mills gauge 
theory at finite temperature, the usual axial gauge quantization, ${\bf A}_0$$=$$0$, 
has been replaced by a modified axial gauge quantization, which also spontaneously breaks the
nonabelian gauge symmetry: $\beta {\bf A}_0 $$=$$ 2\pi (1^8 ;  0^8 ) $ 
\cite{decon,fermionic}. We emphasize that the necessity for a Wilson line background in
the finite temperature quantization arose as a consequence of our insistence on preserving the
finiteness, and perturbative renormalizability, of the zero temperature ground state at
finite temperatures.

\vskip 0.1in
Let us derive from our expression for the string 
free energy the growth with temperature in the low energy field
theory limit, namely, large $\beta$. It will suffice to carry out this check in a single 
sector of the thermal
spectrum, for instance, ${\cal O}_{0}$, with unit winding, but summing over all
integer Matsubara modes, $n$ $\in$ ${\rm Z}$. We emphasize that modular invariance
is not at issue here, since our interest is only in a field theoretic check of the scaling
behavior as a function of $T$ in the leading term of the string level expansion. 
Expanding the Jacobi theta functions in powers of $q{\bar{q}}$ $=$ $e^{-4\pi \tau_2 }$,
and extracting the coefficients of the $O(1)$ term in the level expansion of the zero thermal
winding number sector of ${\cal O}_0$, gives the following leading contribution
from massless target spacetime bosons:
\begin{eqnarray}
 F(\beta)|_{\rm (w=1;n \in {\rm Z} )}  =&& 
     -   2 \cdot 2^8 \cdot  L^{9} (4\pi^2 \alpha^{\prime})^{-5} 
    \int_{- 1/2 }^{ 1/2 } d \tau_1
\int_{{\sqrt{1-\tau_1^2}} }^{\infty}   
      {{d \tau_2 }\over{4\tau_2^2}} \cdot 
  (\tau_2)^{-4}  \cr
  &&\quad \quad \quad 
 \times ~  \sum_{n=-\infty}^{\infty} ~ {\rm exp} \left [ - \pi \tau_2 \left ( {{4\pi^2 \alpha' n^2
    }\over{\beta^2}}  +  {{\beta^2}\over{ 4 \pi^2 \alpha^{\prime} }} \right )   + 2\pi i n \tau_1 \right ]
 \cr
=&& - 2^9 L^{9} T^{10} 
    \sum_{n=-\infty}^{\infty} \left (   1+ {{4\pi^2 \alpha' n^2}\over{\beta^4 }} 
          \right )^{-5}  \int_{- 1/2 }^{1/2 } d \tau_1  {\rm exp}\left [ 2 \pi i n \tau_1 \right ] 
    \Gamma \left ( -5 , {\sqrt{1-\tau_1^2}} \right )   \quad   .
\label{eq:IIidh}
\end{eqnarray}
The leading coefficient of the product of eta functions is $({\bar{q}}^{1/2} q)$ in heterotic
string theory, and we isolate the compensating powes in the $q$ expansion.
Notice that all of the massless target space fermions of the zero temperature vacuum have
acquired a tree-level mass that is {\em linear} in temperature, another self-consistency 
check for the low energy finite temperature gauge theory.
The degeneracy at the massless level accounts for both the 64 bosonic states in the 
spin 2 gravity multiplet, as well as the 8.496 bosonic states in the spin 1 Yang-Mills multiplet.
As an aside, as mentioned before 
in our discussion of the type II string level expansion in section 2, each degree of freedom
in the spin 1 multiplet contributes with half as much weight as those in the spin 2 multiplet.
The reason is that the level expansion counts target space degeneracies in terms of 
{\em equivalent 2d free bosonic oscillator modes}.  The leading $T^{10}$ dependence in 
the free energy arises from the Matsubara spectrum;  the subleading
temperature dependence comes from the background of stringy winding modes. 

\subsection{Thermal Duality and High Temperature Scaling Behavior}

\vskip 0.1in The $T^2$ high temperature scaling behavior of the free energy
can be inferred more elegantly by 
application of the T-duality transformation linking the thermal ground states of 
the two heterotic string theories. That argument proceeds as follows.
In the closed bosonic string theory, the generating
functional for connected one-loop vacuum string graphs is
invariant under the thermal self-duality transformation: $W(T)$
$=$ $W(T^2_c/T)$, at the string scale, $T_c$ $=$ $1/2 \pi
\alpha^{\prime 1/2}$. It was pointed out by Polchinski
\cite{polchinskibook} that we can infer the following thermal duality
relation which holds for both the Helmholtz free energy, $F(T)$
$=$ $-T \cdot W(T)$, and the vacuum energy density, $\rho(T)$ $=$
$-T \cdot W(T)/V$ of the closed bosonic string ensemble:
\begin{equation}
F(T)  = {{T^2}\over{T_C^2}} F({{T_C^2}\over{T}}) , \quad \quad
\rho(T) = {{T^2}\over{T^2_C }} \rho({{T_C^2}\over{T}}) \quad .
\label{eq:thermi}
\end{equation}
In the case of the heterotic string, we will show that the thermal
duality relation instead relates, respectively, the free energies
of the $E_8$$\times$$E_8$ and ${\rm Spin}(32)/{\rm Z_2}$ theories.
Since we deal with a supersymmetric string theory, it is
convenient to restrict ourselves to the contributions to the
vacuum energy density from target space {\em bosonic} degrees of
freedom alone:
\begin{equation}
F(T)_{E_8\times E_8}  = {{T^2}\over{T_C^2}}
F({{T_C^2}\over{T}})_{{\rm Spin}(32)/{\rm Z}_2} , \quad \quad
\rho(T)_{E_8 \times E_8}  = {{T^2}\over{T^2_C }}
\rho({{T_C^2}\over{T}})_{\rm Spin (32)/Z_2} \quad .
\label{eq:thermih}
\end{equation}
Consider the high temperature limit of this expression:
\begin{equation}
\lim_{T \to \infty } \rho(T)_{E_8 \times E_8} = \lim_{T \to
\infty} {{T^2}\over{T^2_C }} \rho({{T_C^2}\over{T}})_{\rm Spin
(32)/Z_2}
 =  \lim_{(T_C^2/T) \to 0} {{T^2}\over{T^2_C }}
\rho({{T_C^2}\over{T}})_{\rm Spin(32)/Z_2}
 =  {{T^2}\over{T^2_C }} \rho(0)_{\rm Spin(32)/Z_2}
\quad , \label{eq:thermasy}
\end{equation}
where $\rho(0)$ is the contribution to the cosmological constant,
or vacuum energy density, at zero temperature from target space
bosonic degrees of freedom alone. Note that it is finite. Thus, at
high temperatures, the contribution to the free energy of either
heterotic ensemble from target space bosonic degrees of freedom
alone grows only as fast as $T^2$. In other words, the growth in
the number of target spacetime bosonic degrees of
freedom at high temperature in
the heterotic string ensemble is only as fast as in a {\em
two-dimensional} field theory. This is significantly slower than
the $T^{10}$ growth of the high temperature degrees of freedom
expected in the ten-dimensional low energy field theory. Notice
that the thermal duality transformation interchanges the thermal
winding and thermal momentum modes of the corresponding 
heterotic ensembles. This is the reason for the simple $T^2$ 
scaling behavior found in this duality relation. 

\vskip 0.1in
Thus, at high temperatures far above the string scale, the leading behavior is a 
significantly {\em slower} $T^2$ growth in the closed superstring free energy. Such
a $T^2$ growth was first conjectured by Atick and Witten in 1989 \cite{aw}. Subsequently, 
it was shown to hold for the free energy of the closed bosonic string ensemble by
Polchinski \cite{polchinskibook} as a
direct consequence of its thermal self-duality property.

\subsection{High Temperature Expansion of Free Energy}

\vskip 0.1in As a final bit of closed string 
methodology, we will now show how the modular integrals
appearing in the result for the one-loop heterotic string 
vacuum amplitude can be carried out in closed form 
by the procedure of 
term-by-term
integration. The result will be a power series 
expansion for the one-loop 
free energy of the
canonical ensemble expressed {\em as an exact function of temperature}. 

\vskip 0.1in
We begin with a Poisson resummation on the infinite 
sum over winding modes, in order
 to put our expression for $F(T)_{\rm het} $ in a
form suitable for a high temperature expansion valid in the regime $\beta$ $\to$ $0$.
Denoting the coefficients in the level expansion by $b_m^{({\rm het})}$,
the result takes the form:
\begin{eqnarray}
 F_{ \rm  het} =&& -
  L^{9}  (4\pi^2 \alpha^{\prime})^{-5}
 \int_{- 1/2 }^{ 1/2 } d \tau_1
\int_{{\sqrt{1-\tau_1^2}} }^{\infty}  {{d \tau_2 
}\over{4\tau_2^{2}}} 
  \tau_2^{-4}   \sum_{n=-\infty}^{\infty}  \sum_{w=-\infty}^{\infty}  \sum_{m=0}^{\infty}  \cr 
  && \quad \quad \times ~  b_m^{\rm (het)} ~  {\rm exp} \left [- 4 \pi m \tau_2  \right ] ~
  {\rm exp} \left [ - \pi \tau_2 \left ( {{4\pi^2 \alpha' n^2
    }\over{\beta^2}}    + {{ w^2 \beta^2}\over{4\pi^2 \alpha'}} \right )  + 2 \pi i n w  \tau_1 \right ]  \cr
    =&& -
    L^{9} (4\pi^2 \alpha^{\prime})^{-5} \cdot \beta^{-1} 
 \int_{- 1/2 }^{ 1/2 } d \tau_1
\int_{{\sqrt{1-\tau_1^2}} }^{\infty}  {{d \tau_2 
}\over{4\tau_2^{2}}} 
  \tau_2^{-4}   \left ( {{ \tau_2 } \over{ 4\pi^2 \alpha^{\prime} }} \right )^{-1/2}  
  \sum_{n=-\infty}^{\infty} \sum_{p=-\infty}^{\infty}  \sum_{m=0}^{\infty}  \cr 
  && \quad \quad \times ~  b_m^{\rm (het) } ~  {\rm exp} \left [- 4 \pi m \tau_2  \right ] ~
    {\rm exp} \left [ - \pi \tau_2 \left ( {{4\pi^2 \alpha' n^2
    }\over{\beta^2}} \right )    - \pi (p - n \tau_1)^ 2 {{4 \pi^2 \alpha^{\prime} }\over{\tau_2 \beta^2 }}  \right ] 
       \quad .
\label{eq:typeIIfoss}
\end{eqnarray}
Notice that the three integer-valued infinite summations 
in the expression for $F_{\rm het} (\beta)$ given here 
account for, respectively, closed string mass level, thermal windings, and thermal momenta. 
It is
convenient to split the range of integration for the modular variable $\tau_2$ as 
follows:
\begin{equation}
\int_{{\sqrt{1-\tau_1^2}} }^{\infty}  d \tau_2 ~   \equiv 
    \int_{0 }^{\infty}  d \tau_2  ~-~ 
      \int_0^{{\sqrt{1-\tau_1^2}} }  d \tau_2 
    \quad , 
\label{eq:intger}
\end{equation}
where in the second term we must take care to express the integrand in a 
form that is manifestly bounded within the specified domain of integration. 
We will find that the former can be recognized as the Bessel function,
$K_{\nu}(y)$, with a power series representation in the
argument $y$:
\begin{eqnarray}
&& \int_{0 }^{\infty} d x ~  x^{\nu -1} ~ e^{-{{\delta}\over{x}} - \gamma x } 
 ~=~  2 \left ( {{\delta}\over{\gamma}} \right )^{\nu/2} K_{\nu}  \left ( 2 {\sqrt{\delta\gamma}}  \right ) , 
 \quad {\rm Re}~ \delta > 0 , ~ {\rm Re}~ \gamma > 0 \cr 
 && \quad {\rm where} ~ \delta \equiv  {{4 \pi^3 \alpha^{\prime} }\over{ \beta^2 }} (p - n \tau_1)^ 2 , 
       ~ \gamma \equiv 4 \pi m + 4 \pi^3 \alpha' n^2/\beta^2   , ~ {\rm and} ~ \nu \equiv  -11/2  \quad .
\label{eq:int1}
\end{eqnarray}
The latter can be expressed in terms of the standard Whittaker functions,
$W_{-{{\nu+1}\over{2}}, {{\nu}\over{2}}} (y)$,  with integral
representation:
\begin{eqnarray}
&& \int_{0 }^{u} d x ~ x^{\nu -1} ~ e^{- \delta / x } 
 ~=~  \delta^{(\nu -1)/2 } u^{ (1+\nu )/2 } e^{-\delta/2 u} 
                  W_{- {{\nu+1}\over{2}} , {{\nu}\over{2}} }  \left ( \delta / u \right )  \cr
                 && \quad {\rm where}   ~ 
                     \delta \equiv   {{4 \pi^3 \alpha^{\prime} }\over{ \beta^2 }} (p - n \tau_1)^ 2, 
                     ~ u \equiv {\sqrt{1-\tau_1^2}} , ~ {\rm and} ~ \nu \equiv -{{11}\over{2}} + k , ~ k \in {\rm Z}^+ 
                     \quad , 
%GR3.471
\label{eq:int12}
\end{eqnarray}                    
following substitution of the Taylor expansion of the function ${\rm exp}[\gamma \tau_2]$
in the integrand. Note that, with this substitution, we 
can verify that the integrand is always bounded in the region $|\tau_2| < 1$, 
term-by-term in the Taylor expansion. 
Substituting in the expression for the free energy given in Eq.\ (\ref{eq:typeIIfoss})
we obtain the result:
\begin{eqnarray}
 F_{ \rm  het} =&& -
  L^{9} (4\pi^2 \alpha^{\prime})^{-5}
 \int_{- 1/2 }^{ 1/2 } d \tau_1
\int_{{\sqrt{1-\tau_1^2}} }^{\infty}  {{d \tau_2 
}\over{4\tau_2^{2}}} 
  \tau_2^{-4}   \sum_{n=-\infty}^{\infty}  \sum_{w=-\infty}^{\infty}  \sum_{m=0}^{\infty}  \cr 
  && \quad \quad \times ~  b_m^{\rm (het) } ~  {\rm exp} \left [- 4 \pi m \tau_2  \right ] ~
  {\rm exp} \left [ - \pi \tau_2 \left ( {{4\pi^2 \alpha' n^2
    }\over{\beta^2}}    + {{ w^2 \beta^2}\over{4\pi^2 \alpha'}} \right )  + 2 \pi i n w  \tau_1 \right ]  \cr
    =&& -
     L^{9} (4\pi^2 \alpha^{\prime})^{-5} \cdot \beta^{-1} 
 \int_{- 1/2 }^{ 1/2 } d \tau_1
\int_{{\sqrt{1-\tau_1^2}} }^{\infty}  {{d \tau_2 
}\over{4\tau_2^{2}}} 
  \tau_2^{-4}   \left ( {{ \tau_2 } \over{ 4\pi^2 \alpha^{\prime} }} \right )^{-1/2}  
  \sum_{n=-\infty}^{\infty} \sum_{p=-\infty}^{\infty}  \sum_{m=0}^{\infty}  \cr 
  && \quad \quad \times ~  b_m^{\rm het } ~  {\rm exp} \left [- 4 \pi m \tau_2  \right ] ~
    {\rm exp} \left [ - \pi \tau_2 \left ( {{4\pi^2 \alpha' n^2
    }\over{\beta^2}} \right )    - \pi (p - n \tau_1)^2 {{4 \pi^2 \alpha^{\prime} }\over{\tau_2 \beta^2 }}  \right ] \cr
=&& -  {{1}\over{4}} L^9 (4\pi^2 \alpha^{\prime})^{-5} \cdot \beta^{-1} 
 \int_{- 1/2 }^{ 1/2 } d \tau_1 
  \sum_{n=-\infty}^{\infty} \sum_{p=-\infty}^{\infty}  \sum_{m=0}^{\infty} ~  b_m^{\rm (het) }  
   (p-n \tau_1 )^{-11/2} 
    \cr 
  &&\quad \quad \quad \quad \quad ~ \times 
   \left [  n^2 + {{m\beta^2}\over{\pi^2 \alpha^{\prime} }}  \right ]^{11/4} ~ K_{-11/2} 
        \left (   {{8\pi^2 \alpha^{\prime} }\over{\beta^2}} (p-n \tau_1 ) \left ( n^2 + {{m\beta^2}\over{\pi^2 \alpha^{\prime} }} \right )^{1/2}   \right )  
        \cr
        && \quad~ ~ - ~  {{1}\over{4}}  L^{9} (4\pi^2 \alpha^{\prime})^{-5}  \cdot \beta^{-1} 
 \int_{- 1/2 }^{ 1/2 } d \tau_1 
  \sum_{n=-\infty}^{\infty} \sum_{p=-\infty}^{\infty}  \sum_{m=0}^{\infty}
        \sum_{k=0}^{\infty}  {{ b_m^{\rm (het) } }\over{k!}}   \cr
            && \quad \quad \quad \times ~ \left [ {{4\pi^3 \alpha^{\prime} }\over{\beta^2}} 
            \left ( n^2 + {{m\beta^2}\over{\pi^2 \alpha^{\prime} }} \right ) \right  ]^{k}   \left [  {{4 \pi^3 \alpha^{\prime} }\over{ \beta^2 }} (p - n \tau_1)^ 2 \right ]^{(k-13/2)/2} ~  (1-\tau_1^2)^{(k-9/2)/4}  
            \cr 
            &&\quad\quad \quad \quad \quad \times ~ {\rm exp} \left [ - {{2\pi^3 \alpha^{\prime} }\over{\beta^2}} {{(p-n \tau_1 )^2 }\over{ {\sqrt{1-\tau_1^2}}  }}  \right ]
   ~ W_{-(k-9/2)/2 , (k-11/2)/2} \left (  {{4\pi^3 \alpha^{\prime} }\over{\beta^2}} {{ (p-n \tau_1 )^2 }\over{ {\sqrt{1-\tau_1^2}}  }} 
   \right ) 
        \quad . \cr &&
\label{eq:typeIr}
\end{eqnarray}
Expressing, in turn, the Bessel, Whittaker, and exponential functions as convergent power 
series expansions in their respective arguments, enables all of the $\tau_1$ dependence in the
integrand to be extracted in the form of an algebraic power series. The term-by-term integration
over $\tau_1$ can then be carried out explicitly. Note the useful relations $W_{\lambda , \mu}(y)$$=$$W_{\lambda, -\mu}(y)$,  $K_{\nu}(y)$$=$$K_{-\nu}(y)$. We will make use of
the following power series formulae:
\begin{eqnarray}
%GR8.468
&&K_{11/2} (y) = {\sqrt{ {{\pi}\over{2y}} }} e^{- y} 
    \sum_{r=0}^{5} {{(5+r)!}\over{r! (5-r)! }}  (2y)^{-r}  , 
 \quad {\rm and}  ~ y \equiv   {{8\pi^2 \alpha^{\prime} }\over{\beta^2}} (p-n \tau_1 )
      \left ( n^2 + {{m\beta^2}\over{\pi^2 \alpha^{\prime} }} \right )^{1/2}   \quad .
\label{eq:bess}
\end{eqnarray}
Notice that, term-by-term, the $\tau_1$ dependence in this series is remarkably simple. Also,
notice that the second subscript, $\mu$, in the Whittaker function 
is such that $2\mu$ is half-integer. We can use a 
functional relation, and the power series representation:
\begin{eqnarray}
%GR9.232, 9.231
&&W_{\lambda , \mu} (z) ~=~  {{\Gamma(-2\mu)}\over{\Gamma( \half - \mu - \lambda )}}
M_{\lambda , \mu} (z) + {{\Gamma ( 2\mu)}\over{\Gamma ( \half + \mu - \lambda )}} M_{\lambda , -\mu } 
(z) \cr
&& M_{ - \mu + \half  , -\mu} (z) = - \left ({{ 1 }\over{2\mu}}\right )  z^{\half -\mu} e^{- \half z} \cr
&&  {\rm where} ~  z \equiv {{4\pi^3 \alpha^{\prime} }\over{\beta^2}} {{ (p-n \tau_1 ) }\over{ {\sqrt{1-\tau_1^2}}  }}   ,   ~ \lambda = - (k-9/2)/2  ,  ~ 
   \mu \equiv  (k- 11/2)/2  \quad .
\label{eq:wht}
\end{eqnarray}
Prior to $\tau_1$ integration, it is convenient to 
include the Taylor expansion of the exponential,  $e^{-z}$, as well as
the powers of binomials, $(p-n\tau_1)$,  $(1-\tau_1^2)^{1/2}$,  accompanying the Whittaker
function in the expression in Eq.\ (\ref{eq:typeIr}), defining the 
coefficients, $w^{(\pm)}_{(r,s)} (\beta, \alpha^{\prime})$, and numerical 
subscripts, $\sigma_{\pm} (\lambda , \mu)$, 
 as follows:
\begin{eqnarray}
&& (p-n\tau_1)^{(k-13)/2} (1- \tau_1^2)^{(k-9/2)/4}   e^{-\half z } W_{-(k-9/2)/2, (k-11/2)/2 } (z) 
\cr && \quad \equiv 
 ~ \sum_{r=0}^{\infty}  \sum_{s=0}^{\infty}  \tau_1^{r+s+ \sigma_{\pm} } 
        \left [ {{\Gamma(-2\mu)}\over{\Gamma( \half - \mu - \lambda )}} w^{(+)}_{(r,s)}   +  
        {{\Gamma ( 2\mu)}\over{\Gamma ( \half + \mu - \lambda )}}  w_{(r,s)}^{(-)}  \right ]
         \quad   .
\label{eq:whtt}
\end{eqnarray}
A similar expansion holds for the term proportional to $K_{11/2}(y)$, with the coefficient
$\kappa(\alpha^{\prime},\beta)$ defined by the relation,
$y$ $\equiv$ $(p-n\tau_1) \kappa$.
The result for $F_{ \rm  het} $ consequently takes the form:
\begin{eqnarray}
&&F_{ \rm  het }  =  - {{1}\over{4}}  L^{9} (4\pi^2 \alpha^{\prime})^{-5} 
  \sum_{n=-\infty}^{\infty} \sum_{p=-\infty}^{\infty}  \sum_{m=0}^{\infty} ~   b_m^{\rm (het)}
      \left [  n^2 + {{m\beta^2}\over{\pi^2 \alpha^{\prime} }}  \right ]^{11/4}   \cr 
  &&\quad \quad \times 
   \int_{- 1/2 }^{ 1/2 } d \tau_1  (p-n\tau-1)^{11/2} \left [ {\sqrt{ {{\pi}\over{2\kappa (p-n\tau_1)}} }} 
   e^{- \kappa (p-n \tau_1) } 
    \sum_{r=0}^{5} {{(5+r)!}\over{r! (5-r)! }}  (2\kappa(p-n\tau_1))^{-r} \right ]
           \cr
        &&  \quad \quad \quad -  ~  {{1}\over{4}} L^{9} \beta^{-1} (4\pi^2 \alpha^{\prime})^{-5} 
  \sum_{n=-\infty}^{\infty} \sum_{p=-\infty}^{\infty}  \sum_{m=0}^{\infty}
        \sum_{k=0}^{\infty}  {{b_m^{\rm (het)}}\over{k!}}   \left [ 
            n^2 + {{m\beta^2}\over{\pi^2 \alpha^{\prime} }} \right  ]^{k}  
                \left [  {{4 \pi^3 \alpha^{\prime} }\over{ \beta^2 }} \right ]^{(3k-13/2)/2}
 \cr
   && \quad   \times   \int_{- 1/2 }^{ 1/2 } d \tau_1 \left \{           
             \sum_{r=0}^{\infty}  \sum_{s=0}^{\infty}  \tau_1^{r+s+ \sigma_{\pm}} 
        \left [ {{\Gamma(-2\mu)}\over{\Gamma( \half - \mu - \lambda )}} w^{(+)}_{(r,s)}   +  
        {{\Gamma ( 2\mu)}\over{\Gamma ( \half + \mu - \lambda )}}  w_{(r,s)}^{(-)}  \right ]
              \right \}               \quad .
\label{eq:typeIrr}
\end{eqnarray}
With these substitutions, all of the $\tau_1$ integrals appearing in our expression for $F_{\rm het} $
can be evaluated explicitly. Recall that $m$ and $n$ denote heterotic string mass level, and thermal momentum number, respectively.
The infinite summation over $p$ resulted from the Poisson
resummation of thermal winding numbers, necessary to cast the expression for the free
energy in a form suitable
for a high temperature expansion in $\beta$. Thus, following the one-loop modular integrations, 
we obtain an exact high temperature expansion for the one-loop string
free energy as a function of $\beta$, and the degeneracies, 
$b_m^{\rm (het)}$, in the heterotic string mass level expansion. 

\vskip 0.1in Finally, it should be noted that, using the power series 
representation in $\beta$ given above, one could 
demonstrate the analyticity of infinitely many thermodynamic
potentials for the heterotic ensemble. Namely, potentials defined by taking
arbitrarily large number of temperature derivatives in the vicinity of the
self-dual temperature. This is clear evidence that the thermal
duality transition linking the high temperature behavior of one
heterotic ensemble to the low temperature behavior of the other ensemble lies
within the Kosterlitz-Thouless universality class 
\cite{kt,sathia,kogan,nair,bosonic}. 

\section{Free Energy of the Unoriented Type IB Ensemble}

\vskip 0.1in The free
energy of the unoriented open and closed type IB string ensemble
at one-loop order in the string coupling 
receives contributions from surfaces of four different worldsheet
topologies \cite{polchinskibook}: torus, annulus, Mobius
strip, and Klein bottle. Given our analysis of the 
heterotic and type II superstring ensembles in previous sections, we
expect that the Yang-Mills fields will play an essential role in 
enabling discussion 
of an equilibrium canonical ensemble. As explained in section 2, 
pure type II closed string theories have no Yang-Mills gauge fields, but upon
coupling to Dbranes, the full unoriented open and closed string theory
indeed contains timelike Wilson line backgrounds. This is the case
we shall consider in this section. 

\vskip 0.1in 
The torus contribution sums over closed
{\em oriented} worldsheets, and the result therefore follows from the analysis 
of the type IIB superstring described in section 2. 
In particular, there are no tachyon-free thermal ground states
that break target space supersymmetry and, consequently, no 
equilibrium canonical ensemble.  The 
 functional, $W_{\rm tor}(\beta)$, therefore vanishes identically.
Thus, at one-loop order in 
string perturbation theory, the oriented closed string sector will not contribute
to the free energy of the type IB thermal vacuum.

\vskip 0.1in 
Thus, our focus must shift to the open and unoriented string 
sectors of the type IB thermal spectrum. An {\em equilibrium} description of 
type IB string statistical
mechanics in the canonical ensemble requires
 a tachyon-free
thermal spectrum in the full temperature range, in addition to the absence
of massless Ramond-Ramond sector tadpoles.
The vacuum functional at one-loop order in the
presence of $N$ $=$ $2^5$ D9branes can be written in the
general form:
\begin{eqnarray}
W_{\rm IB}(\beta) = && \beta L^9
(4\pi^2 \alpha^{\prime})^{-5} \int_0^{\infty} {{dt}\over{2t}}
   t^{-5}  {\rm Z}_{\rm IB} (\beta) 
\quad ,   \label{eq:freein}
\end{eqnarray}
where the function ${\rm Z}_{\rm IB} (\beta)$ denotes the level
expansion of the type IB unoriented thermal ensemble, with contributions
from annulus, Mobius strip, and Klein bottle.

\vskip 0.1in
Notice that there is no constraint
analogous to modular invariance on the open 
string mass spectrum. In addition, there are no thermal winding modes.
Consider the contribution to the path integral from the $n$th Matsubara mode:
\begin{equation}
e^{- 4\pi^3 \alpha^{\prime}n^2 t  /\beta^2 } ,  \quad \quad n , ~ n+ \half \in {\rm Z} \quad  .
\label{eq:mode}
\end{equation}
Focussing on the $t$$\to$$\infty$ IR limit of the string amplitude at first, at
low temperatures $\beta$ $\to$ $\infty$, we see that the field theoretic modes
with low values of $n$ dominate. But in the opposite $t$ $\to$ $0$ limit, we
must include arbitrarily high values of $n$ in the Matsubara spectrum. 
What about the high temperature limit, with $\beta$ $\to$ $0$? Even in
the $t$ $\to$ $\infty$ limit, modes with arbitrarily high $n$ are now important.
Thus, a thermal duality transformation to the Euclidean T-dual type I$^{\prime}$
ensemble would illuminate the high temperature behavior of the ensemble.
As in the case of the heterotic
string ensemble, the key will lie in introducing a temperature dependent 
Wilson line gauge background in order to achieve the spontaneous breaking of 
supersymmetry without the introduction of thermal tachyons.
We must simultaneously require that ${\rm Z}_{\rm (IB)} (\beta)$ 
preserve the Ramond-Ramond sector massless 
tadpole cancellations necessary for infrared finiteness and perturbative renormalizability
of the finite temperature vacuum. 

\vskip 0.1in 
Recall that the Wilson line introduced in the ${\rm Spin}(32)/{\rm Z}_2$ 
heterotic string breaks the nonabelian gauge group to $SO(16)$$\times$$SO(16)$,
while spontaneously breaking supersymmetry. Let us compactify the type IB SO(32) 
string on a circle of radius $\beta/2\pi$, turning on the temperature dependent Wilson line 
\cite{dlp,dbrane,polchinskibook}:
\begin{equation}
\beta A_0 =  ( 1^8 , 0^8 ) ,  \quad \quad \theta_i = 2\pi  , \quad  i =1, \cdots , 8;
~ \theta_i = 0 , \quad  i=9, \cdots , 16 \quad . 
\label{eq:thetaw}
\end{equation}
This is the SO(16)$\times$SO(16) type IB vacuum, and the 
Wilson line wrapping the $X^0$ coordinate has spontaneously broken supersymmetry
in addition to a partial breaking of 
the nonabelian gauge symmetry in the supersymmetric SO(32) vacuum.
Thus, we wish to find the finite temperature vacuum functional that interpolates
between the tachyon-free and nonsupersymmetric type IB SO(16)$\times$SO(16) 
string at finite $\beta$, and the supersymmetric type IB $SO(32)$ string in the 
limit where the Wilson loop is infinite radius. 

\vskip 0.1in
The Euclidean T-duality transformation, $\beta$ $\to$ 
$\beta^{\prime}$$=$$4\pi^2 \alpha^{\prime} /\beta$, 
maps the configuration of D9branes coincident with an $O9$ plane
to a configuration of a pair of  $O8$ planes, each coincident with 
16 D8branes, at the two endpoints of an interval of length $\beta^{\prime}$.
The thermal modes of the type IB ensemble correspond to a
Matsubara-like frequency spectrum with timelike momentum:
$p_n$$=$$2n\pi/\beta$, where $n$, $n+\half$ $\in$ ${\rm Z}$; for the T-dual
type I$^{\prime}$ ensemble this role is played by the infinite
summations over thermal winding modes.
Note that the D8branes are each paired with their images, so that the gauged background
has a total of $8^2$ oriented stretched strings of length $\beta^{\prime}$.
Supersymmetry is spontaneously broken due to
the presence of the stretched strings, which
contribute with positive thermal tension to the 
vacuum energy of the finite temperature vacuum:
\begin{equation}
{\cal T} (\beta) =  \beta^2 /4 \pi^2 \alpha^{\prime}  \quad ,
\label{eq:wils}
\end{equation}
contributing an overall shift in the masses of the thermal tachyons in the NS sector. 
The GSO projections on worldsheet fermions eliminates all tachyons from the 
physical state spectrum, as in the supersymmetric vacuum.

\vskip 0.1in 
The cancellation of both RR-RR, and NS-NS, sector massless tadpoles, and the absence of
open string tachyons, in a nonsupersymmetric and tachyon-free 9D 
SO(16)$\times$SO(16) type IIB orientifold, was shown by Blum and Dienes in \cite{bd}.
As a consequence, both the cylinder and the Klein bottle amplitude vanish identically in
the 9D vacuum, and the result for the vacuum energy density 
can be written entirely in terms of the contribution from the Mobius strip:
\begin{eqnarray}
{\rm Z}^{\rm NS-NS}_{\rm Mob} (\beta) =&&  \left [ {\cal E}_0 + {\cal E}_{1/2} \right ] \left
({{\Theta_{01}(it;0)\Theta_{10}(it;0)}\over{\eta(it)\Theta_{00}(it)}}
\right )^4  \cr
{\rm Z}^{\rm R-R}_{\rm Mob} (\beta) =&&  \left [ {\cal E}_0 -  {\cal E}_{1/2} \right ] \left
({{\Theta_{01}(it;0)\Theta_{10}(it;0)}\over{\eta(it)\Theta_{00}(it)}}
\right )^4  \quad .
\label{eq:so1m}
\end{eqnarray}
The authors of Ref.\  \cite{bd} have computed the individual contributions from the
Mobius strip and torus amplitudes numerically 
down to values of the radius at which the divergence in
the torus amplitude sets in, as a consequence of the closed 
string (high temperature) 
thermal tachyon. The asymptotic approach to the supersymmetric SO(32)
vacuum at low temperatures is evident in their plot. As explained earlier, we believe 
it is more meaningful to interpret these results as evidence that the worldsheet 
RG flow is {\em towards} the IR 
stable supersymmetric vacuum in the closed string sector: the one-loop vacuum amplitude
computes the tree level mass spectrum, and at this order in string perturbation theory the
supergravity and Yang-Mills sectors have not as yet \lq\lq communicated". Thus, 
supersymmetry
remains unbroken in the gravitational sector, and the nonvanishing one-loop 
vacuum energy density of the type I model comes wholly from the Mobius strip.

\vskip 0.1in We believe the correct physical interpretation of the Blum-Dienes result
is as follows. Unlike the heterotic case, in the type I strings, there are thermal interpolations
between the 10D supersymmetric SO(32) heterotic vacuum and either the 9D nonsupersymmetric
and nontachyonic type IB, or the T-dual type I$^{\prime}$, SO(16)$\times$SO(16) string vacua.
The former has only thermal momentum modes, and the latter only thermal winding modes.
The result for the free energy of the type I$^{\prime}$ ensemble takes the form:
\begin{eqnarray}
F_{\rm I^{\prime} }(\beta) = && -  L^9
(4\pi^2 \alpha^{\prime})^{-5} \int_0^{\infty} {{dt}\over{2t}}
    t^{-5}  [\eta(it)]^{-8} 
   \left [ {{\Theta_{01}(it;0)\Theta_{10}(it;0)}\over{\eta(it)\Theta_{00}(it)}}\right ]^4 \cr
   && \quad\quad\quad\quad\quad\quad \quad \quad \quad \quad \times ~ 
\sum_{w=-\infty}^{\infty} \left \{ e^{-2 \pi \beta^2 w^2 /4\pi^2 \alpha^{\prime} } +  
e^{-2 \pi \beta^2  (w+ \half )^2 ]/4 \pi^2 \alpha^{\prime}}  \right \}
 \quad ,  \label{eq:freeinp}
\end{eqnarray}
or the Euclidean T-dual transformed result for the type IB ensemble:
\begin{eqnarray}
F_{\rm IB }(\beta) = && -  L^9
(4\pi^2 \alpha^{\prime})^{-5} \int_0^{\infty} {{dt}\over{2t}}
   t^{-5}  [\eta(it)]^{-8} 
   \left [ {{\Theta_{01}(it;0)\Theta_{10}(it;0)}\over{\eta(it)\Theta_{00}(it)}}\right ]^4 \cr
   && \quad\quad\quad\quad\quad\quad\quad \quad \quad \quad \times ~ 
\sum_{n=-\infty}^{\infty} \left \{ e^{-8 \pi^3 \alpha^{\prime} n^2 /\beta^2} +  
e^{-8 \pi^3 \alpha^{\prime} (n+ \half )^2 ]/\beta^2}  \right \}
 \quad ,  \label{eq:freeinpb}
\end{eqnarray}
Thus, we have found evidence for a stable equilibrium ensemble of type I unoriented
open and closed strings in the limit of weak string coupling. 

\subsection{The Low Energy Limit of $F_{\rm IB}(T)$}

\vskip 0.1in 
At low temperatures far below the string mass scale,
$\beta$$>>$$\alpha^{\prime 1/2}$,  we expect not to excite any
thermal modes beyond the lowest-lying field theoretic modes in the open
string spectrum.  A useful check of
self-consistency with the low energy finite temperature
field theory limit is to verify the expected $T^{10}$ growth of the free
energy. For simplicity, we will restrict ourselves to the sector of the
thermal spectrum with half-integer moding of Matsubara frequencies.
An analogous expression holds for the sector with integer moding.

\vskip 0.1in We begin with the one-loop contribution to the type
IB string vacuum functional given in Eq.\ (\ref{eq:freeinpb}), 
including both integer and 
half-integer Matsubara frequencies. Expand
 the integrand of the modular integral in powers of
$q$$=$$e^{-2\pi t}$, thus isolating the $t$$\to$$\infty$
asymptotics of the modular integral which is dominated by the
massless modes in the open string spectrum. We have:
\begin{eqnarray}
F_{\rm IB} (\beta) =&& - L^9 (4\pi^2 \alpha^{\prime})^{-5}
 \int_0^{\infty} {{dt}\over{2t}} ~
  t^{-5} 
[\eta(it)]^{-8}  \left [ {{\Theta_{01}(it;0)\Theta_{10}(it;0)}\over{\eta(it)\Theta_{00}(it)}}\right ]^4
\cr
&& \quad\quad\quad \times ~ \sum_{n =-\infty}^{\infty} 
\left [ e^{-8 \pi^3 \alpha^{\prime} n^2 /\beta^2} +  
e^{-8 \pi^3 \alpha^{\prime} (n+ \half )^2 /\beta^2}  \right ] \quad .
 \label{eq:freeItex}
\end{eqnarray}
Notice that at sufficiently low temperatures, an infinity
of Matsubara modes can contribute to the $\beta$ $\to$ $\infty$ limit of 
this expression since the integral on $t$ ranges over
[$0$, $\infty$]. Thus, we must include the full sum over the full
Matsubara spectrum when computing the leading low energy
field theory limit of the string free energy.

\vskip 0.1in
We recognize in our expression the integral
representation of an Euler gamma function with {\em negative}
argument, $\Gamma(-9/2)$, which may be defined by either the
product formula for gamma functions, or by analytic continuation
in the argument of the gamma function. Thus, performing the
explicit integration over $t$ for the leading term in the level 
expansion on $m$, and restricting to the summation over just the
half-integer moded Matsubara spectrum for illustration, gives the result:
\begin{eqnarray}
F_{\rm short} = && - L^9 (4\pi^2 \alpha^{\prime})^{-5}
 \cdot 2 \sum_{n = 0 }^{\infty}  \cdot  \int_0^{\infty} {{dt}\over{2t}} ~
  t^{-5} \cdot 2^4 \cdot e^{- 8
\alpha^{\prime}\pi^3 (n+ \half )^2 t /\beta^2} \cr = && - L^9 (4\pi^2
\alpha^{\prime})^{-5} 2^{4} 
 ~ \left [ \zeta(-10 , \half ) \right ] \left [ (2\pi)^{5} \Gamma \left (-5 \right )
(4\pi^2 \alpha^{\prime})^{5} \beta^{-10} \right ]   \cr
  \equiv&& -  \left [ L^9 (4\pi^2 \alpha^{\prime})^{-5} \right ]
(\beta_0/\beta)^{10}  \quad , \label{eq:freet}
\end{eqnarray}
where $L^9$ denotes the nine-dimensional spatial volume.
$\Gamma(-n)$ would have to be defined by analytic continuation, invoking a
complex integral representation.
The factors within the second pair of square brackets in Eq.\
(\ref{eq:freet}) result from the integral over the world-sheet
modulus, $t$. The modular integration is followed by the infinite
summations on even and odd integers, giving the result within the first pair
of square brackets. The summations have been expressed in terms of
the Riemann zeta function $\zeta (z,q)$:
\begin{equation}
 \sum_{n = 0}^{\infty} (n+ \half )^{-z} \equiv  \zeta (z, \half ) , \quad \quad \quad \zeta(-n  ,\half ) = - 
    {{1 }\over{ (n+1) (n+2) }}  B_{n+2}^{\prime} (x)|_{x=\half}  \quad ,  \label{eq:zeta}
\end{equation}
and $B_{n}(x)$ is a Bernoulli polynomial, $B_n^{\prime}(x)$$=$$nB_{n-1} (x)$.
The parameter, $\beta_0$, has the dimensions of an inverse
temperature, and it characterizes the asymptotic limit of the free
energy for the low energy supersymmetric gauge theory. Recall that
the string vacuum functional is dimensionless. Thus, we have
demonstrated that the one-loop free energy, $F(\beta)$$=$$-W_{\rm
IB} (\beta)/\beta$, grows as $T^{10}$ in the asymptotic low
temperature regime. This is precisely the behavior expected of a
ten-dimensional, finite temperature supersymmetric gauge theory.
Explicitly, we have:
\begin{equation}
\beta_0 = [ 2^{4}  (B_{11}( \half)/11) \Gamma
(-5) ]^{1/10} (4\pi^2 \alpha^{\prime})^{1/2}  \quad .
\label{eq:beta0}
\end{equation}

\subsection{High Temperature Expansion}

\vskip 0.1in Conversely, when we include the contribution from all
of the massive string modes, we find 
a $T^2$ growth in the string free energy at temperatures far
above the string scale. Set $t$$\to$$1/t$ in the expression for the 
level expansion, and expand in powers of
$e^{-\pi/t}$. The ultraviolet asymptotics of
the mass level expansion dominates the high temperature regime, giving the following
leading contribution to
 $F_{\rm IB} (\beta)$:
\begin{eqnarray}
 F_{\rm IB} (\beta) = && -  
L^9 (4\pi^2 \alpha^{\prime})^{-5}
   \int_0^{\infty}{{ dt}\over{2t}}  ~
  {{t^{-1}}\over{
  \eta(i/t)^{8}}} \cr
  \quad && \quad \times ~ 2 ~ \sum_{n=0}^{\infty} 
\left [ {{\Theta_{01}(i/t;0)\Theta_{10}(i/t;0)}\over{\eta(i/t)\Theta_{00}(i/t)}}\right ]^4
 e^{-8 \alpha^{\prime}\pi^3 (n+ \half )^2 t /\beta^2} \cr
 =&& -  L^9 (4\pi^2 \alpha^{\prime})^{-5}  
 \sum_{n = 0}^{\infty}  \int_0^{\infty} dt ~
   t^{-2}\left [  2^4 + O(e^{-\pi/ t})
\right ] e^{-8  \alpha^{\prime}\pi^3 (n+ \half )^2 t /\beta^2} \cr = && - L^9
(4\pi^2 \alpha^{\prime})^{-5}  \left [ 2^4 \zeta(- 2  , \half )
 \right ] \left [ \Gamma(-1) (2\pi) (4\pi^2
\alpha^{\prime}) \beta^{-2} \right ] \cr
 \equiv&&  - L^9 (4\pi^2
\alpha^{\prime})^{-5} (\beta_1/ \beta)^2  \quad .
\label{eq:freeht}
\end{eqnarray}
Thus, we find that the free
energy at one-loop order in {\em every} perturbative string
ensemble, whether open or closed, bosonic or supersymmetric, 
grows as $T^2$ at temperatures far above the string scale, 
a conjecture originally made by Atick and Witten
\cite{aw} for closed string theories alone, and later shown
to be a consequence of thermal self-duality in the closed bosonic
string theory by Polchinski
\cite{polchinskibook}.

\vskip 0.1in 
Including all of the
corrections to the ultraviolet asymptotic limit of the open
string mass spectrum, we find
that the 
term-by-term integration over the modulus, $t$, can be expressed in
terms of the Bessel function $K_{1}
(x)$. The free energy takes the form:
\begin{eqnarray}
F_{\rm IB} (\beta) = && - L^9 (4\pi^2 \alpha^{\prime})^{-5}
 \cdot 2 \sum_{n = 0 }^{\infty}  \int_0^{\infty} {{dt}\over{2t}} 
t^{-1} \left [ 2^4 + \sum_{m=1}^{\infty} b_m^{\rm (IB)} 
e^{-2\pi m/t} \right ] e^{- 8 \alpha^{\prime}\pi^3 (n+ \half )^2 t /\beta^2} \cr
=&& - L^9 (4\pi^2 \alpha^{\prime})^{-5} \left [ (\beta_1/\beta)^2 
+  \sum_{n = 0 }^{\infty}  \int_0^{\infty} dt
t^{-2} \sum_{m=1}^{\infty} b_m^{\rm (IB)}  e^{-2\pi m/t}  e^{-8
\alpha^{\prime} \pi^3 (n+ \half )^2 t /\beta^2} \right ]  \cr &&
\label{eq:freetws}
\end{eqnarray}
where the $b_m^{\rm (IB)}$ are the coefficients in the
mass level expansion for the type IB thermal mass spectrum, obtained by 
expanding in powers of
$q$$=$$e^{-2\pi/ t}$:
\begin{equation}
{\cal Z}_{\rm IB} (i/t) \equiv [\eta(i/t)]^{-8} 
\left [ {{\Theta_{01}(i/t;0)\Theta_{10}(i/t;0)}\over{\eta(i/t)\Theta_{00}(i/t)}}\right ]^4
\equiv \sum_{m=0}^{\infty} b_m^{\rm (IB)}  e^{-2\pi m/t}
  \quad . \label{eq:expi}
\end{equation}
Performing the integration over $t$ results in the expression:
\begin{eqnarray}
F_{\rm IB} (\beta) = &&  - L^9 (4\pi^2 \alpha^{\prime})^{-5}
  (\beta_1/\beta)^2 \cr
  && \quad\quad 
 - L^9 (4\pi^2 \alpha^{\prime})^{-5} 
\sum_{m=1}^{\infty} b_m^{\rm (IB)} \sum_{n = 1}^{\infty}  \left [ 2\pi (4\pi^2  
\alpha^{\prime } )(n+ \half )^2/\beta^2) \right ] \cdot  \left [  2 ({{x}\over{2}} )^{-1}
K_1 (x) \right ]^{-1}    \quad , \cr
 && \label{eq:fretwst}
\end{eqnarray}
where $\beta^2 x^2 $$=$$ 64 \pi^4 \alpha^{\prime} m
(n+\half)^2 $ identifies the argument of the Bessel function. 
We emphasize that this result for the type IB free energy is 
valid at all temperatures, and includes all of
the massive string modes, as well as the full half-integer moded Matsubara
frequency spectrum. A similar expression can be derived for the
integer moded frequency spectrum. The factor
in square brackets in the first equality arises from the modular
integration. The infinite summation over thermal mode numbers 
therefore yields $\zeta( -1 , \half )$. The large $x$, or high temperature,
asymptotics of the Bessel function 
indicates clearly that all of the corrections to the leading $T^2$ growth of
the string free energy 
are exponentially suppressed. 
The asymptotic expression for the one-loop free energy 
therefore takes the form:
\begin{equation}
%GR8.339
F_{\rm IB} (T)/ L^9 (4\pi^2 \alpha^{\prime})^{-5}  =  -  \beta_1^2 T^2 - \sum_{m=1}^{\infty} \sum_{n
= 1}^{\infty} A_{m} T e^{-\beta_2(m,n) T} \quad .
\label{eq:freeIB}
\end{equation}
where the leading $T^2$ growth is found to be corrected by a
negligible $T e^{-T}$ dependence, at high temperatures. Recall that
$\beta_1$
takes the form:
\begin{eqnarray}
\beta_1^2 =&& - \left [ (2\pi) 2^4 \Gamma(-1) B_3 (\half)/3 \right ] \beta_C^2 
 \quad ,  \label{eq:params}
\end{eqnarray}
where we have substituted for $\zeta(-2 , \half )$. As before, $\Gamma(-n)$
has to be defined by analytic continuation.
Notice the minus sign in the coefficient,
reversing the sign of the free energy, and clarifying that the dominant
UV contribution
is from target space fermionic modes. 

\vskip 0.1in It is helpful to check the corresponding scaling
behavior as a function of temperature
 for the first few thermodynamic potentials. The internal
energy of the canonical ensemble,
$U $ $=$ $ - \left ( \partial W/ \partial \beta  \right )_V $, displays
the same scaling behavior as the free energy.
The entropy is given by the expression:
\begin{equation}
S = - \beta^2 \left ( {{\partial F}\over{\partial \beta }}\right
)_V = - \beta^2 \left [ \beta^{-2} W(\beta) - \beta^{-1} \left (
{{ \partial W}\over{\partial \beta }} \right )_V  \right ] \quad ,
\label{eq:entropy1n}
\end{equation}
and we infer that it scales at high temperatures as $\beta^{-1}$.
Finally, since the specific heat at constant volume is given by:
\begin{equation}
C_V = - \beta \left ( {{\partial S} \over{\partial \beta }} \right
)_V \quad , \label{eq:sppairf}
\end{equation}
we infer that it also scales as $\beta^{-1}$ at high temperatures.
The corresponding scaling behavior at {\em low} temperatures, in
agreement with the expected result for the low energy gauge theory
limit can be extracted, as in the previous section, by considering
instead the $t$$\to$$\infty$ asymptotics of these expressions.

\section{Conclusions}

\vskip 0.1in We have shown that all of the perturbatively renormalizable
and anomaly-free supersymmetric
string theories with Yang-Mills gauge fields: heterotic, 
type I, and type I$^{\prime}$ admit stable and tachyon-free 
finite temperature ground states in which we can formulate an equilibrium
statistical mechanics of strings in the weak coupling limit. Preceeding attempts \cite{bow,ky,bt,follow,aw,micro,bv} to
formulate an equilibrium statistical mechanics with a tachyon-free
supersymmetric string canonical ensemble have failed for a variety
of reasons; a detailed account appears in
section 2 of hep-th/0409301v1. These works
do not correctly incorporate the Euclidean T-duality
transformations linking the thermal vacua of the supersymmetric
string theories in pairs. Neither do they meet the infrared consistency
conditions we have required in our analysis, matching self-consistently
with the low energy finite temperature field theoretic physics. In addition,
string theoretic consistency conditions, such as modular invariance, have often
been violated. The results in this paper demonstrate irrefutably 
that the widespread misconception
that the canonical ensemble of superstrings necessarily breaks down
beyond a limiting (Hagedorn) temperature is simply wrong. Finally,
in the case of the pure type IIA and type IIB thermal ground states, we
show that it is not possible to eliminate low temperature tachyons 
from the thermal spectrum while meeting the constraints from modular
invariance. This fact also precludes interpretation of thermal winding mode
tachyons as indication of a string scale \lq\lq Hagedorn" phase transition
in the type II theories:
the canonical ensemble cannot be defined {\em in the full
temperature range, starting at $T$$=$$0$}. 

\vskip 0.1in  For each of the six superstring
ensembles, we have shown that the growth of the free energy
at high temperatures far
above the string scale is only as fast as that in a 2d
 quantum field theory: $F(T)$ $\simeq$ $T^2$. This behavior exemplifies
the equivalence of perturbatively renormalizable superstring theories
to the 2d gauge theory of diffeomorphism and Weyl invariances.
We have simultaneously verified that the low energy field theoretic limit
of our expressions for the string free energy in each case recovers
the expected $T^{10}$ growth characteristic of a 10D quantum field
theory, at low temperatures where only the lowest-lying massless field
theory modes contribute. These results follow directly from an
 explicit evaluation
of the one-loop modular integrals.  The $T^2$ growth of the 
high temperature string free energy 
was originally conjectured by Atick and Witten in \cite{aw}. We have
shown that the $T^2$ scaling
can also be inferred from the thermal duality relations
linking the thermal ground states of the superstring ensembles in
pairs. This is in precise analogy with the $T^2$ scaling of the free energy of the 
closed bosonic string as a consequence of its thermal self-duality, as shown
by Polchinski in \cite{polchinskibook}.
In fact, we can infer the existence of a duality phase
transition in the Kosterlitz-Thouless universality class mapping
the finite temperature ground state of the $E_8$$\times$$E_8$
heterotic string to its ${\rm Spin}(32)/{\rm Z}_2$ Euclidean
T-dual, analogous to the self-duality phase
transition observed in the closed bosonic
string ensemble \cite{bosonic}. The Kosterlitz-Thouless universality
class is characterized by an infinite hierarchy of thermodynamic
potentials displaying analyticity at the critical temperature, in this
case, $T_C$ $=$ $1/2\pi \alpha^{\prime 1/2}$ \cite{kt, sathia,kogan,bosonic}.

\vskip 0.1in We will close with mention of an important insight
that applies more broadly to the development of a fundamental, and
nonperturbative, formulation for 
String/M Theory.
We remind the reader that perturbative string theory
as formulated in the worldsheet formalism is inherently background
dependent: the \lq\lq heat-bath" representing the embedding target
space of fixed spatial volume and fixed inverse temperature is forced
upon us, together with any external background fields
characterizing the target spacetime geometry. Thus, we are ordinarily
restricted to the canonical ensemble of statistical mechanics. We
should caution the reader that while an immense, and largely
conjectural, literature exists on proposals for microcanonical
ensembles of weakly-coupled strings \cite{ea,bt,long,bv,micro},
the conceptual basis of these treatments is full of holes. Some of
the pitfalls have been described in \cite{svet,aw}. One
could argue that, strictly speaking, the microcanonical ensemble
is what is called for when discussing quantum cosmology, or the
statistical mechanics of the Universe \cite{hawk}: the Universe
is, by definition, an isolated closed system, and it is
meaningless to invoke the canonical ensemble of the \lq\lq
fundamental" degrees of freedom. However, there remain many simpler
questions in both early Universe cosmology, and in finite temperature
gauge theories, that are 
approachable within the framework of the
canonical ensemble. Limited use of the microcanonical ensemble
under certain assumptions is also possible, but the constraints
imposed by the Jeans instability, and by thermal back reaction, 
need to be kept in mind.

\vspace{0.1in}\noindent{\bf Acknowledgements:}
I would like to thank Joe Polchinski for urging me to be
precise in applying the thermal (Euclidean T-duality) 
transformations linking the thermal ground states of the
six superstring theories. Although this comment
came in May 2001, as clarified in Footnote 2, the idea of
pursuing the significance of thermal duality also comes
from his previous work \cite{polchinskibook}. I would like to
acknowledge Hassan 
Firouzjahi for correcting a misleading typo in presentations
of the type II NS-NS tachyon spectrum prior to Aug 2004. I 
also thank an anonymous reader, as
well as Keith Dienes, Mike Lennek, and Lubos Motl, for pointing
out an error in my presentations of modular invariance
in the type II analysis prior to June 2005. Both corrections
reinforce my conclusion in \cite{fermionic} of the impossibility 
of a type II
canonical ensemble in the absence of a Yang-Mills gauge sector. 
This research
has been supported in part by the National Science Foundation, 
the Aspen Center for Physics, 
and the Kavli Institute for Theoretical Physics.

\end{document}